\documentclass[pre,superscriptaddress,showpacs,nofootinbib,
  floatfix,preprint,preprintnumbers]{revtex4-1}

\usepackage[utf8x]{inputenc}
\usepackage{amsfonts}
\usepackage{amsmath}
\usepackage{amssymb}
\usepackage{graphicx}
\usepackage{hyperref}
\usepackage{bm}

\DeclareGraphicsExtensions{.pdf}

\newcommand{\cch}[1]{\left[#1\right]}

\newcommand{\prt}[1]{\left(#1\right)}
\newcommand{\aver}[1]{\left\langle #1 \right\rangle}
\newcommand{\soma}[3]{\ensuremath\displaystyle\sum_{#1}^{#2} #3}

\hypersetup{colorlinks=true, citecolor=blue,
  pagecolor=blue,
  urlcolor=black,
  pdfcreator={pdflatex},
}

\begin{document}
\title{Critical points, phase transitions and water-like anomalies for
  an isotropic two length scale potential with increasing attractive well}

\author{L. Pinheiro }
\author{A. P. Furlan}
\affiliation{\footnotesize Instituto de Física, Univeridade Federal do Rio Grande do
  Sul, Caixa Postal 15051, CEP 91501-570, Porto Alegre, RS,
  Brazil.}

\author{L. B. Krott}
\address{\footnotesize Centro Ararangu\'a, Universidade Federal de Santa Catarina, Rua
Pedro Jo\~ao Pereira, 150, CEP 88905-120, Ararangu\'a, SC, Brazil}

\author{A. Diehl}
\affiliation{\footnotesize Departamento de Física,Instituto de Física e Matemática,
Universidade Federal de Pelotas, Caixa Postal 354,
CEP 96010-900 Pelotas, RS, Brazil}

\author{M. C. Barbosa}
\affiliation{\footnotesize Instituto de Física, Univeridade Federal do Rio Grande do
  Sul, Caixa Postal 15051, CEP 91501-570, Porto Alegre, RS,
  Brazil.}

\begin{abstract}
Molecular Dynamic and Monte Carlo studies are performed in 
a family of 
core-softened (CS) potential, composed by two length scales: a
repulsive shoulder at short distances and the another a variable scale, that
can be repulsive or strongly attractive depending on the
parameters used. 
The density, diffusion and structural anomalous regions 
in the pressure versus temperature phase diagram shrink in 
pressure as the system becomes more attractive.
The liquid-liquid transition appears as a consequence
of the non monotonic behavior of the density versus
pressure isotherms with the increase of the attraction well.
We found that the 
 liquid-gas phase transition is Ising-like for all
the CS potentials and its
critical temperature increases with the increase of the attraction.
No Ising-like behavior for the liquid-liquid phase transition
was detected in the Monte
Carlo simulations 
what might be due to the presence of stable solid phases.

\textit{Keywords:} water anomalies, phase transitions, core-softened potentials
\end{abstract}
\maketitle

%%%%%%%%%%%%%%%%%%%%%%%%%%%%%%%%%%%%%%%%%%%%%%%%%%%%%%%%%%%%%%%%%%%%%
\section{Introduction}
\label{sec:introduction}
%%%%%%%%%%%%%%%%%%%%%%%%%%%%%%%%%%%%%%%%%%%%%%%%%%%%%%%%%%%%%%%%%%%%%

The description of a
single component system as particles interacting via  a 
core-softened (CS) two-body
potentials has being used as viable strategy to understand
the mechanism behind universal phenomena in anomalous liquids. 
These potentials exhibit a
repulsive core with a softening region with a shoulder or a ramp~\cite{lo07}.
These 
models originate from 
the  desire to construct a simple two-body isotropic 
potential capable of 
the density~\cite{Wa64,Ke75,An76} and
diffusion~\cite{An76,Pr87,spce}  anomalies present in water.
Another motivation for these studies is the  acknowledged possibility that
some single component systems display coexistence between two
 different liquid phases~\cite{Po92,Fr10,
Po92,Fr02a,Fr07,Fr03,Ku08}. The use of two length scales
potentials seems to be an interesting tool for finding
the connection between the presence of thermodynamic
and dynamic anomalies and the possibility of the presence
of two liquid phase. 

Complementary  to the thermodynamic
and dynamic anomalies, water also shows an unusual behavior
in its structure. While for normal liquids the
system becomes more structured with the increase 
of the  density, water shows a maximum. Such behavior
can be characterized by translational order parameter $t$
\cite{Sh02,Er01,Er03} that exhibits a region  in which $t$ decreases under
compression. The entropy also shows a very peculiar behavior. 
The excess entropy $S_{ex}$, defined as the difference between the
entropy $S$ of the liquid and the ideal gas, at same density and
temperature~\cite{Ru06b,Ag10,Ag11,Ol06a,Ol06b,Ol07,Ol10a,Er06,Mi06a,
Ya07}, becomes a great tool in the investigation of liquid-state
anomalies~\cite{Ba11}. The region where $(\partial S_{ex}/\partial
\rho > 0)$ on isothermal compression corresponds to an anomaly in excess
entropy, indicating an existence of distinct forms of local ordering,
for a high density limit, where particles are found closer to each other,
and a low density region, with large average distance between particles.

The use of core-softened potentials
to reveal the origin of these anomalies
becomes even more interesting 
because the anomalies mentioned above are not exclusive of
water. Studies have shown that Te\cite{Th76}, Ga,
Bi~\cite{Handbook}, S~\cite{Sa67,Ke83}, Ge$_{\mbox{\tiny 15}}$
Te$_{\mbox{\tiny 85}}$~\cite{Ts91}, BeF$_2$~\cite{An00,Ru06b,Ag07a,
Ag07b}, silica~\cite{An00,Sh02,Ru06b,Po97} and silicon~\cite{Ta97}
present water-like anomalies. 

 The CS potentials show a variety of 
shapes. They can be ramp-like
~\cite{He70,St72,Ja98,Ja99a,Ja99b,Ja01a,Ku05} or continuous shoulder-like~
\cite{Ca03,Ca05,Wi02,Wi06,Ol06a,Ol06b,Ol07,Ol08a,Ol08b,Ba09,Ol09,Si10}. Even
though these work show the presence of the anomalies and
in some cases the existence of the second critical point but
in others the two liquid phases are not 
present~\cite{Fo08,Ol06b,Ol07,Ol08a,Ol08b}, the 
limit in which the presence of the anomalies is related 
to the existence of a second critical point is not clear. In
this paper we employ a family of CS potentials spanning from
purely repulsive to a very attractive case and analyze 
the behavior of the anomalies, liquid-liquid and liquid-gas critical
point indicating the

This paper is organized as follow: in Sec. \ref{sec:the model} we
introduce the model; in Sec.
\ref{sec:simulation} the methods and the
simulation details are described; in Sec. \ref{sec:results} the
results are presented; and finally, in Sec. \ref{sec:conclusion}, the conclusions
are given.

%%%%%%%%%%%%%%%%%%%%%%%%%%%%%%%%%%%%%%%%%%%%%%%%%%%%%%%%%%%%%%%%%%%%%
\section{The model}
\label{sec:the model}
%%%%%%%%%%%%%%%%%%%%%%%%%%%%%%%%%%%%%%%%%%%%%%%%%%%%%%%%%%%%%%%%%%%%%%

The fluid is modeled by spherical particles with diameter
$\sigma$ and mass $m$, that interact through
a three dimensional two length scales potential given by
%%%%%%%%%%%%%%%%%%%%%%%%%%%%%%%%%%%%%%%%%%%%%%%%%%%%%%%%%%%%%%%%%%%%%%
\begin{equation}
  \label{eq:potential}
  \frac{U(r_{ij})}{\epsilon} = \epsilon^{\prime}\cch{
    \prt{\frac{\sigma}{r_{ij}}}^{12} - \prt{\frac{\sigma}{r_{ij}}}^6}
  + \sum_{i=1}^{k}\frac{B_i}{B_i^2+(r_{ij}-C_i)^2},
\end{equation}
%%%%%%%%%%%%%%%%%%%%%%%%%%%%%%%%%%%%%%%%%%%%%%%%%%%%%%%%%%%%%%%%%%%%%%
where $r_{ij}=|{\bm r}_i-{\bm r}_j|$ is the distance between two
particles $i$ and $j$. The potential is composed by a standard
12-6 Lennard-Jones (LJ) potential~\cite{Al89} followed by a
sum of $k$ Lorentzian distributions centered in $C_i$ and with
amplitude $1/B_i$. This composition of the two functions provides a
repulsive shoulder at short distances and an attractive global minimum
at long distances, depending on the set of parameters used.

\begin{figure}[!htb]
  \includegraphics[scale=0.5]{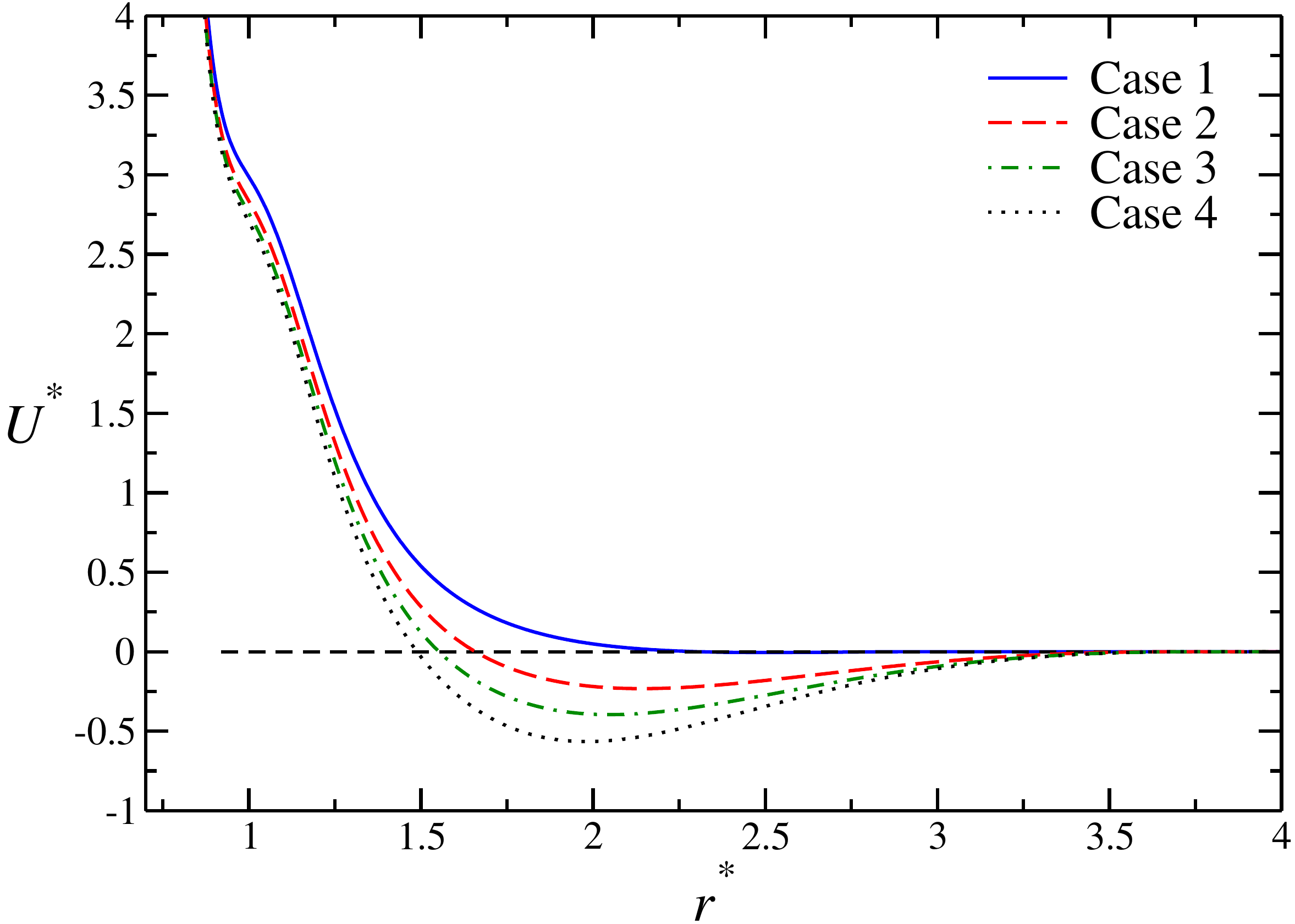}
  \caption{All the potentials studied here. For the Case 1 there is
    just a repulsive shoulder at $r^*\sim 1.0$. For the cases 2, 3 and 4
    the repulsive shoulder is also maintained at $r^*\sim 1$,
    but the attractive part is at $r^*=2.18$, $2.10$ and $2.06$,
    respectively. The black dashed line just represent the level zero
    of potential.% In this sense, in inside picture, we plot
    %$r^*F\prt{r^*}\times r^*$, for the case 5, to demonstrate such
    %behavior.
    }
  \label{fig:potentials}
\end{figure}

The set of parameters were chosen in order to provide different
interaction scales. The idea is to have a purely repulsive 
case in which no liquid-liquid or liquid-gas transitions would
be present. In addition, different energy 
attractive wells were chosen so the liquid-gas
and liquid-liquid transitions  would 
appear.
The potentials resulting from the choice of the parameters are shown
in figure~\ref{fig:potentials}
%%%%%%%%%%%%%%%%%%%%%%%%%%%%%%%%%%%%%%%%%%%%%%%%%%%%%%%%%%%%%%%%%%%%%%
%%%%%%%%%%%%%%%%%%%%%%%%%%%%%%%%%%%%%%%%%%%%%%%%%%%%%%%%%%%%%%%%%%%%%%
The set of parameters that were chosen for each case is shown in
table~\ref{table1}.
%%%%%%%%%%%%%%%%%%%%%%%%%%%%%%%%%%%%%%%%%%%%%%%%%%%%%%%%%%%%%%%%%%%%%%
\begin{table}[!htb]
  \begin{tabular}{p{3.5cm}|p{3.5cm}|p{1.5cm}}
  \hline\hline
  \ Parameters values                   & \ Parameters values     & \
  \ \ Cases \\ \hline
  \ \ \ \ \ \ $\epsilon^{\prime} = 0.60$& \ \ \ \ \ $k=3$         & \\
  \ \ \ \ \ $B_1 = 0.30$                & \ \ \ \ \ $C_1 = 1.00$  & \\
  \ \ \ $B_2 = \begin{cases} -1.0 \\
                         -1.2 \\
                         -1.5 \\
                         -3.0
             \end{cases}$               & \ \ \ \ \ $C_2 = 1.80$ &
             $\begin{cases} \mbox{Case 4} \\ \mbox{Case 3} \\
               \mbox{Case 2} \\ \mbox{Case 1} \end{cases}$\\
   \ \ \ \ \ $B_3 = 2.00$               & \ \ \ \ \ $C_3 = 3.00$ & \\
%   \ \ \ \ \ $B_1=0.23$                 &  \ \ \ \ \ \ \ same & \ \
%\ Case 5 \\
   \hline\hline
 \end{tabular}
 \caption{Parameters of the potentials studied.}
 \label{table1}
\end{table}
%%%%%%%%%%%%%%%%%%%%%%%%%%%%%%%%%%%%%%%%%%%%%%%%%%%%%%%%%%%%%%%%%%%%%%

\begin{figure}
  \includegraphics[scale=0.7]{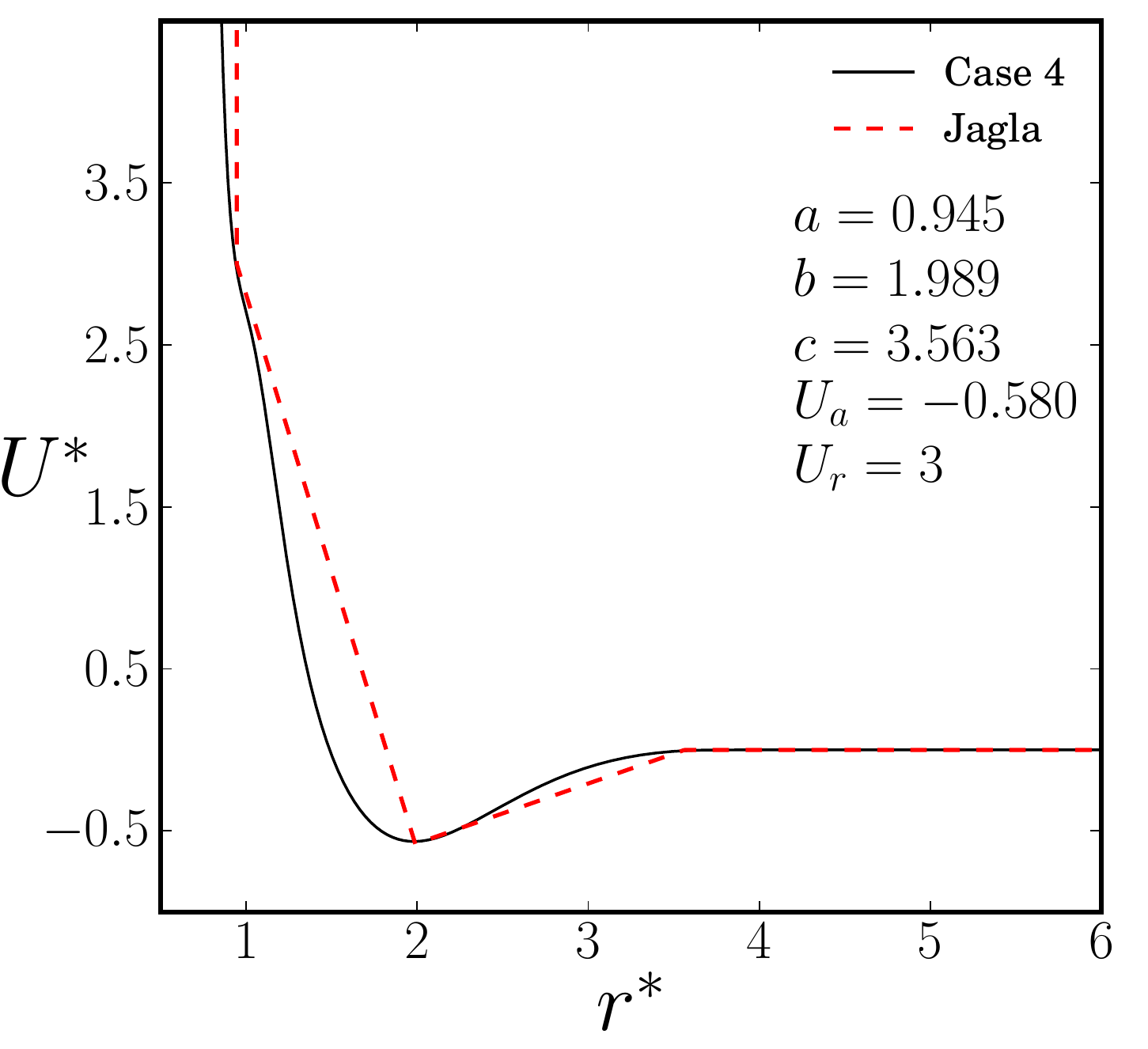}
  \caption{Equivalence between the potential defined by Xu {\it et al.}
    ~\cite{Xu06} and our continuous potential. The interpretation of
    parameters can be seen by Xu {\it et al.}~\cite{Xu06}.}
  \label{fig:comp_lor+jagla}
\end{figure}

Our potential as illustrated in the figure~\ref{fig:comp_lor+jagla}
was constructed to follow
the two length scales   Jagla's ramp  potential~\cite{Ja04}
but with an smooth shape. In our parameterization
the attractive part of the potential was increased
so we can test the effect not only of the continuous forces
but also of the depth of the attraction.
%%%%%%%%%%%%%%%%%%%%%%%%%%%%%%%%%%%%%%%%%%%%%%%%%%%%%%%%%%%%%%%%%%%%%%
%%%%%%%%%%%%%%%%%%%%%%%%%%%%%%%%%%%%%%%%%%%%%%%%%%%%%%%%%%%%%%%%%%%%%%

%%%%%%%%%%%%%%%%%%%%%%%%%%%%%%%%%%%%%%%%%%%%%%%%%%%%%%%%%%%%%%%%%%%%%%
\section{Simulation details}
\label{sec:simulation}
%%%%%%%%%%%%%%%%%%%%%%%%%%%%%%%%%%%%%%%%%%%%%%%%%%%%%%%%%%%%%%%%%%%%%%

Two different simulations techniques were employed:
Molecular Dynamics (MD) and Grand-Canonical Monte Carlo (GCMC)
simulations.
%%%%%%%%%%%%%%%%%%%%%%%%%%%%%%%%%%%%%%%%%%%%%%%%%%%%%%%%%%%%%%%%%%%%%%
\subsection{Molecular Dynamics in the $NVT$ Ensemble }
%%%%%%%%%%%%%%%%%%%%%%%%%%%%%%%%%%%%%%%%%%%%%%%%%%%%%%%%%%%%%%%%%%%%%%
The systems were studied using MD simulations with $512$ particles in
a cubic box, with the standard periodic boundary conditions in all
directions. The simulations were performed in the $NVT$ ensemble,
with the Nose-Hoover~\cite{Ho85,Ho86} thermostat with coupling 
parameter $Q = 2$ to keep
the temperature fixed. The
particle-particle interaction was considered until a characteristic cutoff
radius $r_c=3.7$, and the potential was shifted in
order to provide $U=0$ at $r_c$. The initial configurations of the
systems were chosen as liquid structures. The equilibrium state was
reached after $5 \times 10^5$ steps, followed by $8 \times 10^5$
simulation steps for a production. For the integration of the 
motion equations we have used
the Velocity-Verlet method~\cite{Al89}, with time step
$\Delta t=0.001$ in LJ reduced units. The average of the physical
quantities were obtained using $50$ uncorrelated samples. The thermodynamic
stability of the system was checked by analyzing the dependence of the
pressure on namely and by the behavior of the energy after
the equilibration.

The structural properties of the fluid were obtained by inspection of the
behavior of the translation order parameter~\cite{Sh02,Er01,Er03},
defined as
%%%%%%%%%%%%%%%%%%%%%%%%%%%%%%%%%%%%%
\begin{equation}
  \label{eq:trans_ord_param}
  t \equiv \int_0^{\xi_c}\Big\vert g(\xi)-1\Big\vert d\xi, \quad
  \mbox{where}   \quad\left\{
    \begin{array}{ccc}
      \xi   & = & r\rho^{1/3}, \\
      \xi_c & = & r_c\rho^{1/3},\\
    \end{array}
  \right.
\end{equation}
%%%%%%%%%%%%%%%%%%%%%%%%%%%%%%%%%%%%%
where $r\rho^{1/3}$ represents the average number of particles at a
given distance $r$ and $g(\xi)$ is the radial distribution function.
The radial distribution function is given by 
%%%%%%%%%%%%%%%%%%%%%%%%%%%%%%%%%%%%%
\begin{equation}
  \label{eq:g_of_r}
  g(r)=\frac{V}{N^2}\aver{\soma{i=1}{N}\soma{j=1,j\neq i}{N}
    \delta\prt{{\bm r}-{\bm r}_{ij}}}, \quad r_{ij}=
  \left| \bm{r}_i\prt{\tau}-\bm{r}_j\prt{\tau} \right| .
\end{equation}
%%%%%%%%%%%%%%%%%%%%%%%%%%%%%%%%%%%%%
where ${\bm r}_{i}$ and ${\bm r}_{j}$ are the coordinates of particles $i$ and
$j$ at time $\tau$, $V$ and $N$ are the volume and number of
particles respectively and $\aver{\ldots}$ denotes an average over all
particles.

The dynamical behavior was obtained through the diffusion coefficient
 $D$, related to the mean square displacement (MSD) from Einstein's
relation,
%%%%%%%%%%%%%%%%%%%%%%%%%%%%%%%%%%%%%
\begin{equation}
  \label{difusao_lateral}
  D^*=\lim_{\tau\to\infty}
  \dfrac{\aver{\Delta r{(\tau)^2}}}{6\tau},
  \quad \aver{\Delta r\prt{\tau}^2}=\aver{\cch{{\bm r}_i
      \prt{\tau_0+\tau}-{\bm r}_i(\tau_0)}^2},
\end{equation}
%%%%%%%%%%%%%%%%%%%%%%%%%%%%%%%%%%%%%
where $\Delta r$ represent the distance traveled by a particle between
two steps of integration of the equation of motion.

The excess
entropy, defined as the difference between the entropy of the
real fluid and an ideal gas at the same temperature and density was
also computed. It
can be given by its two body contribution $s_{e}$,
%%%%%%%%%%%%%%%%%%%%%%%%%%%%%%%%%%%%%
\begin{equation}
 s_e \sim s_2 = -2\pi \rho \int_{0}^{\infty} \left[ g(r)\ln g(r)-
   g(r)+1\right]r^2 dr.
\label{eq:s2}
\end{equation}
%%%%%%%%%%%%%%%%%%%%%%%%%%%%%%%%%%%%%

%%%%%%%%%%%%%%%%%%%%%%%%%%%%%%%%%%%%%
\subsection{$\mu VT$ ensemble (GCMC)}
%%%%%%%%%%%%%%%%%%%%%%%%%%%%%%%%%%%%%
In addition to the MD simulations, in 
order to understand the phase behavior of the gas-liquid
transition with the increase of 
the attractive well Grand Canonical
 Monte Carlo (GCMC) simulations~\cite{Fi69} were employed.
The use of the GCMC for this analysis allow us 
to identify the phase coexistence and the stability of 
the different phases.
 In the standard GCMC method the variables chemical potential $\mu$,
volume $V$ and temperature $T$ are fixed, allowing that the total
number of particles $N$ and energy $U$ fluctuate around of 
a mean value. Particles
can be inserted with probability
%%%%%%%%%%%%%%%%%%%%%%%%%%%%%%%%%%%%%
\begin{equation}
 {\cal P}^{\rm acc}(N\to N+1)=\mbox{min}\left[ 1,\;\frac{V}
     {\Lambda^3 (N+1)}e^{\beta(\mu-\Delta U)}
 \right]\;,
\end{equation}
removed with probability
\begin{equation}
 {\cal P}^{\rm acc}(N\rightarrow N-1)=\mbox{min}\left[1,\frac{\Lambda^3 N}{V}
 e^{-\beta(\mu+\Delta U)}
 \right]\;,
\end{equation}
%%%%%%%%%%%%%%%%%%%%%%%%%%%%%%%%%%%%%
and displaced from a initial position $R_i$ to a final position $R_f$
with standard Metropolis~\cite{Me53} probability
%%%%%%%%%%%%%%%%%%%%%%%%%%%%%%%%%%%%%
\begin{equation}
  {\cal P}^{\rm acc}(R_i\to R_f)=\mbox{min}
  \left[ 1,\;e^{-\beta \Delta U}\right]\;.
\end{equation}
%%%%%%%%%%%%%%%%%%%%%%%%%%%%%%%%%%%%%
Here $\Lambda$ is the de Broglie wavelength, $\beta=1/k_BT$ is the inverse
of the thermal energy and $\Delta U$ is the energy difference of the
system, resulting from insertion, removing or displacement movement.

In addition to the GCMC simulations the Hyper-Parallel
Tempering Monte Carlo (HPTMC)~\cite{de99} method  was
performed. The two complementary approaches are 
relevant to avoid metastable state. The idea is to use the
configurations of high temperature to explore the region below the
critical temperature, obtaining good statistics for the gas-liquid
coexistence. The HPTMC is composed of two steps: in the first step,
$N_r$ replicas of the system, at different thermodynamic states, are
simulated in parallel using the standard GCMC procedure. Each
different thermodynamic state is characterized by values of
$\{U_i,N_i,\mu_i,T_i\}$ in the range $\{T_1,\ldots,T_{N_r}\}$ and
$\{\mu_1,\ldots,\mu_{N_r}\}$. In the second part, arbitrary pairs of
replicas have their configurations exchanged with
probability~\cite{de99,Al89}
%%%%%%%%%%%%%%%%%%%%%%%%%%%%%%%%%%%%%%%%%
\begin{equation}
    {\cal P}^{\rm acc}(i\longleftrightarrow j)=\mbox{min}\left[ 1,\;
    \exp \{(\beta_j\mu_j-\beta_i\mu_i)
    (N_i-N_j)-(\beta_i-\beta_j)[U_i-U_j]\}\right].
  \label{eq:parallel}
\end{equation}
%%%%%%%%%%%%%%%%%%%%%%%%%%%%%%%%%%%%%%%%%
A variable number of replicas depending on the case
studied was adopted. In the GCMC simulations  $5\times 10^7$ MC steps to
equilibration and $10^8$ MC steps to data production were employed. The 
simulations
were performed for four box sizes: $L^{\ast}=$ 10, 12, 15 e 18. For
simplicity, only the probability $P(x)$ for the biggest
$L^{\ast}$ size was used.

The critical properties were obtained using the histogram
reweighting~\cite{Fe88} method. The histogram reweighting is a
technique that allow us to obtain thermodynamic averages of a specific state
of system from the trajectory of the other state of system. To this
end, multiple histograms in number of particles $N$ and energy $U$ are
combined, for a region with overlap between the histograms~\cite{Wi06}.
Thus, the probability $\wp (N,U;\mu,\beta)$ of
observe $N$ particles with energy $U$ is given by
%%%%%%%%%%%%%%%%%%%%%%%%%%%%%%%%%%%%%%%%%%%%%%%%%%%%%%%%%%%%%%%%%%%%%%
\begin{equation}
\wp(N,U;\mu,\beta) = \frac{\displaystyle \sum_{i=1}^R
  f_i(N,U)\exp[-\beta U + \beta \mu N]}{\displaystyle \sum_{i=1}^R
K_i\exp[-\beta_i U + \beta_i\mu_i N - C_i]}\;.
\label{prob_tot}
\end{equation}
%%%%%%%%%%%%%%%%%%%%%%%%%%%%%%%%%%%%%%%%%%%%%%%%%%%%%%%%%%%%%%%%%%%%%%
In this equation $K_i$ is the total number of 
observations for a particular run, $f_{i}(N,U)$
is the absolute number of observations of $N$ particles with energy $U$
for a particular run,
and the constants $C_i$ are the ``weights'', which are obtained by the
equation
%%%%%%%%%%%%%%%%%%%%%%%%
\begin{equation}
\exp[C_i] = \sum_E\sum_N\,\wp (N,U;\mu_i,\beta_i)\; .
\label{weights}
\end{equation}
%%%%%%%%%%%%%%%%%%%%%%%%

The absolute pressure was then obtained using the difference between the
``weights'' of two different thermodynamic states~\cite{Fe88}. If the
states are characterized by $(\mu_1,V,\beta_1)$ and
$(\mu_2,V,\beta_2)$, the pressure is obtained by
%%%%%%%%%%%%%%%%%%%%%%%%
\begin{equation}
C_2 - C_1 = \ln \frac{\Xi(\mu_2,V,\beta_2)}{\Xi(\mu_1,V,\beta_1)}=
\beta_2p_2V-\beta_1p_1V\;.
\end{equation}
%%%%%%%%%%%%%%%%%%%%%%%%
Hence, to estimate the
pressure for a specific run, a pressure for reference was
used. For instance, at
low densities we employed the value given
by the ideal gas pressure  $p=\rho k_BT$.

All physical quantities are shown in reduced units~\cite{Al89} as
%%%%%%%%%%%%%%%%%%%%%%%%%%%%%%%%%%%%%%%%%%%%
\begin{eqnarray}
r^* &=& \frac{r}{\sigma} \nonumber \\
U^* &=& \frac{U}{\epsilon} \nonumber \\
\tau^* &=& \frac{(\epsilon/m)^{1/2}}{\sigma} \tau \nonumber \\
T^* &=& \frac{k_B}{\epsilon}T \nonumber \\
p^* &=& \frac{ \sigma ^3 }{\epsilon}p \nonumber \\
\rho^* &=& \sigma ^3\rho \nonumber \\
D^* &=& \frac{(m/\epsilon)^{1/2}}{\sigma}D \nonumber \\
c_V^*&=& \dfrac{c_V}{k_B} \;\; .
\end{eqnarray}
%%%%%%%%%%%%%%%%%%%%%%%%%%%%%%%%%%%%%%%%%%%%%%%%%%%%%%%%%%%%%%%%%%%%%%

%%%%%%%%%%%%%%%%%%%%%%%%%%%%%%%%%%%%%%%%%%%%%%%%%%%%%%%%%%%%%%%%%%%%%%
\section{Results}
\label{sec:results}
%%%%%%%%%%%%%%%%%%%%%%%%%%%%%%%%%%%%%%%%%%%%%%%%%%%%%%%%%%%%%%%%%%%%%%

First, we explore the effects of the attractive energy
of the pair potential in
the presence of 
the density anomaly and in 
the existence of a liquid-liquid critical. 
%%%%%%%%%%%%%%%%%%%%%%%%%%%%%%%%%%%%%%%%%%%%%%%%%%%%%%%%%%%%%%%%%%%%%%
 \begin{figure}[!htb]
  \includegraphics[scale=.4]{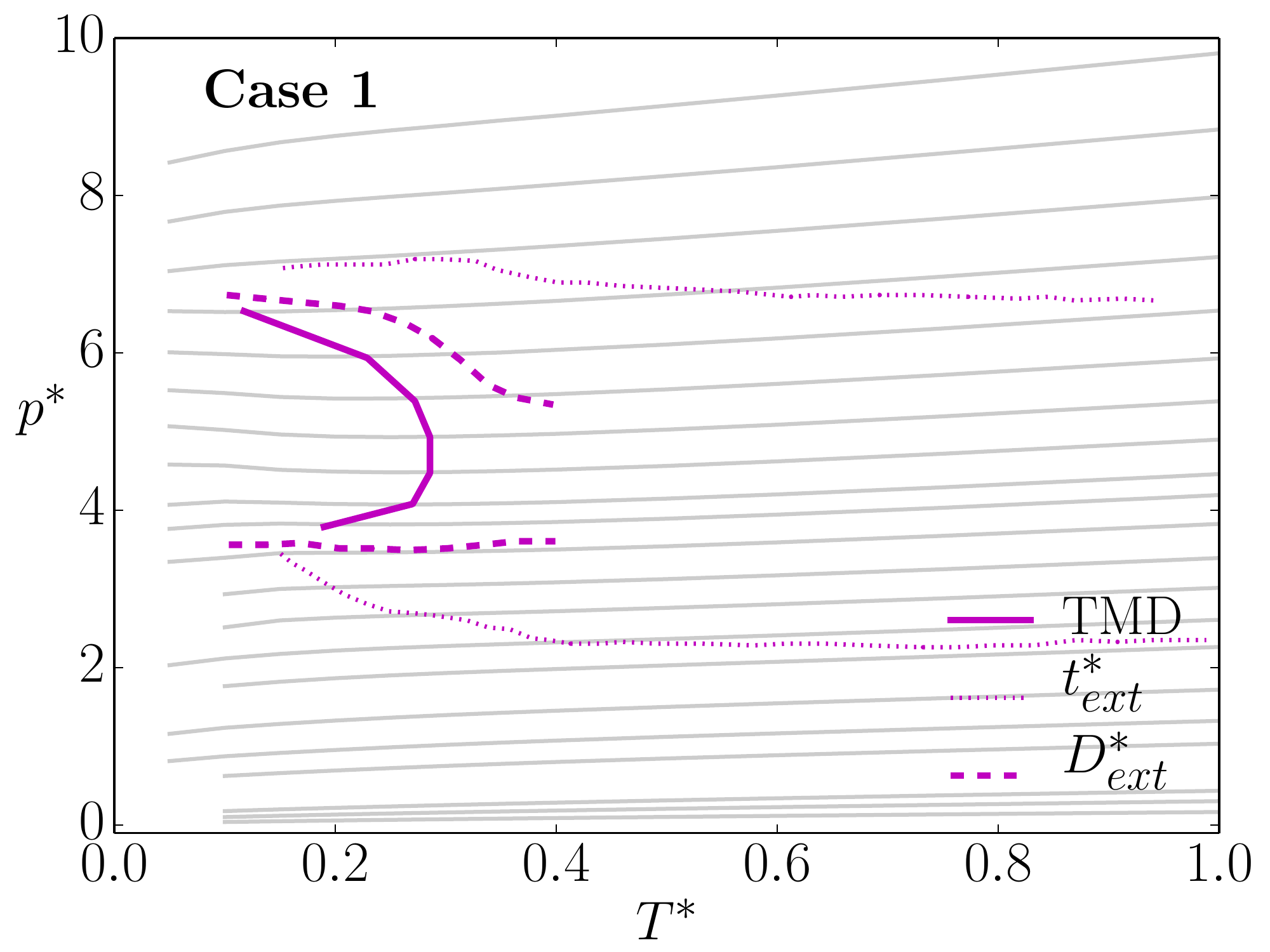}
  \includegraphics[scale=.4]{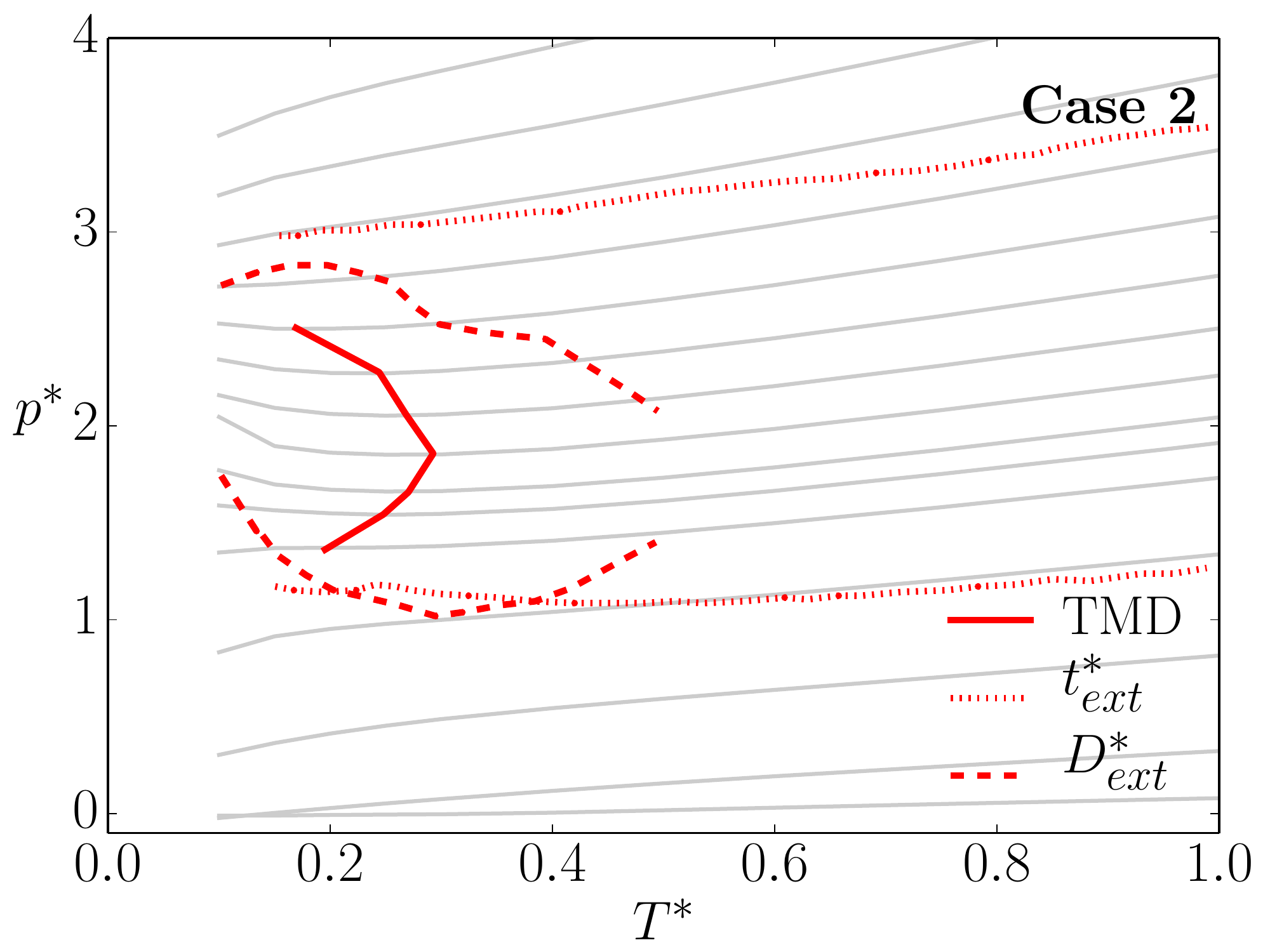}

  \includegraphics[scale=.4]{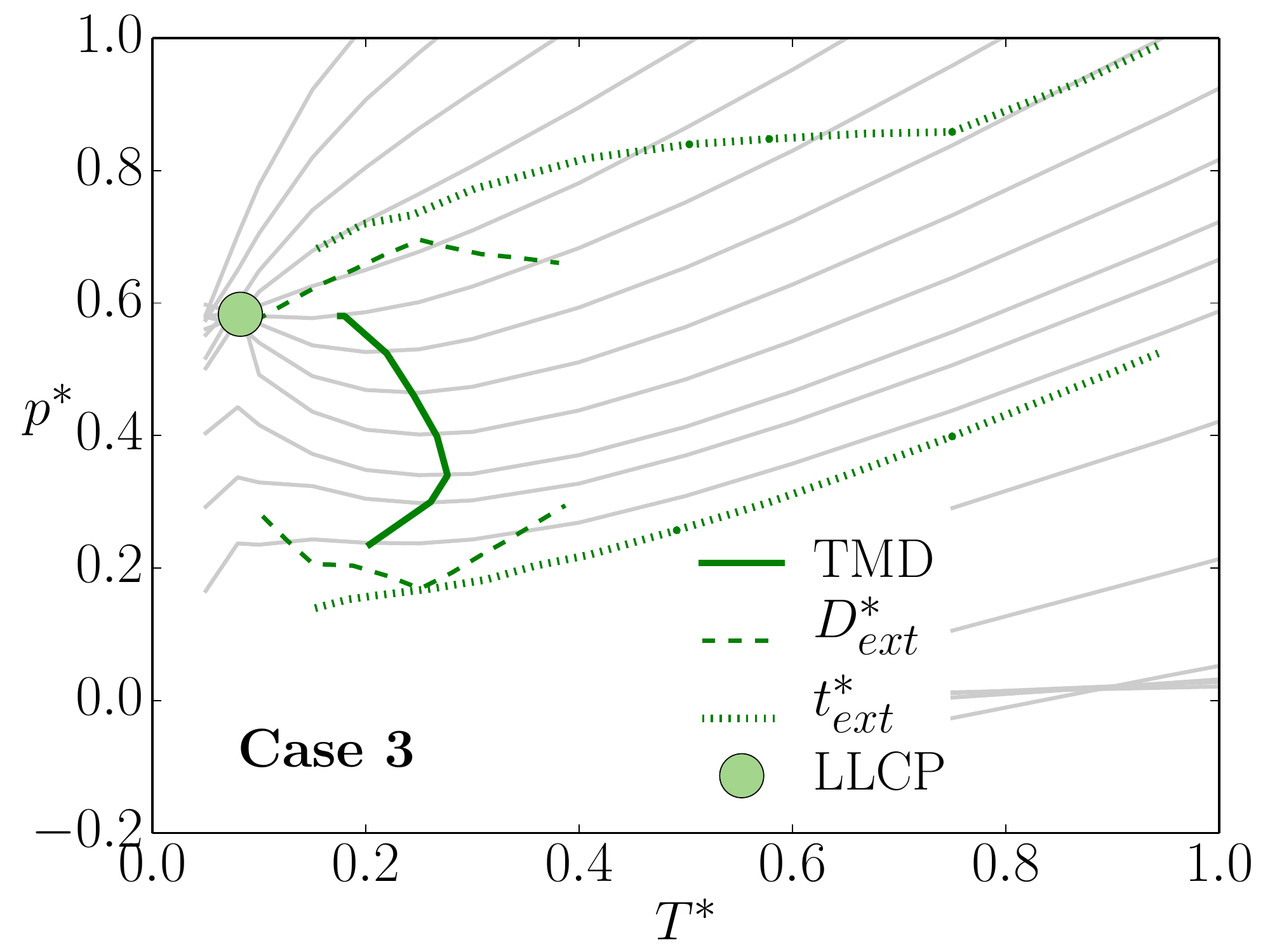}
\includegraphics[scale=.4]{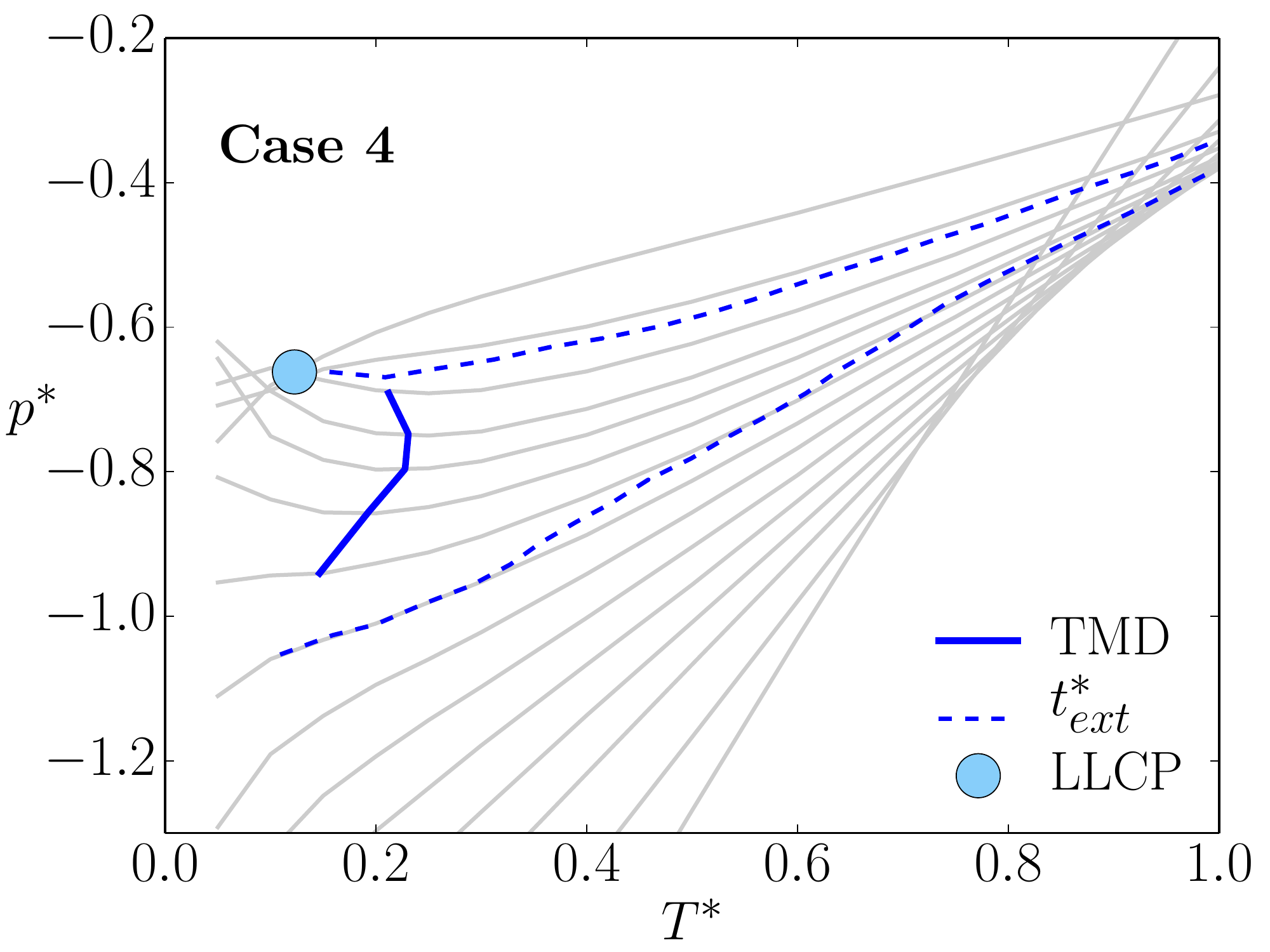}

  \caption{Pressure versus temperature phase diagram for all the 
    potentials studied. The gray lines are the $\rho^*_{1}=0.0284,\dots ,0.652$, 
    $\rho^*_{2}=0.046,\ldots ,0.652$,
    $\rho^*_{3}=0.046,\ldots ,0.81$ and $\rho^*_{4}=0.30,\ldots ,0.81$
    are isochores, the solid lines represent
    the TMD (temperature of maximum density), the dashed lines are the extrema
    in diffusion coefficient and the dotted lines are extrema in
    translational order parameter. The circles represent the liquid-liquid
    critical points for the cases 3 ($p^*_c=0.5831$, $T^*_C=0.0824$) and 
    4 ($p^*_C=-0.6620$, $T^*_C=0.1227$).}
  \label{fig:all_pxyt}
\end{figure}
%%%%%%%%%%%%%%%%%%%%%%%%%%%%%%%%%%%%%%%%%%%%%%%%%%%%%%%%%%%%%%%%%%%%%%

%%%%%%%%%%%%%%%%%%%%%%%%%%%%%%%%%%%%%%%%%%%%%%%%%%%%%%%%%%%%%%%%%%%%%%
 \begin{figure}[!htb]
 \includegraphics[scale=.5]{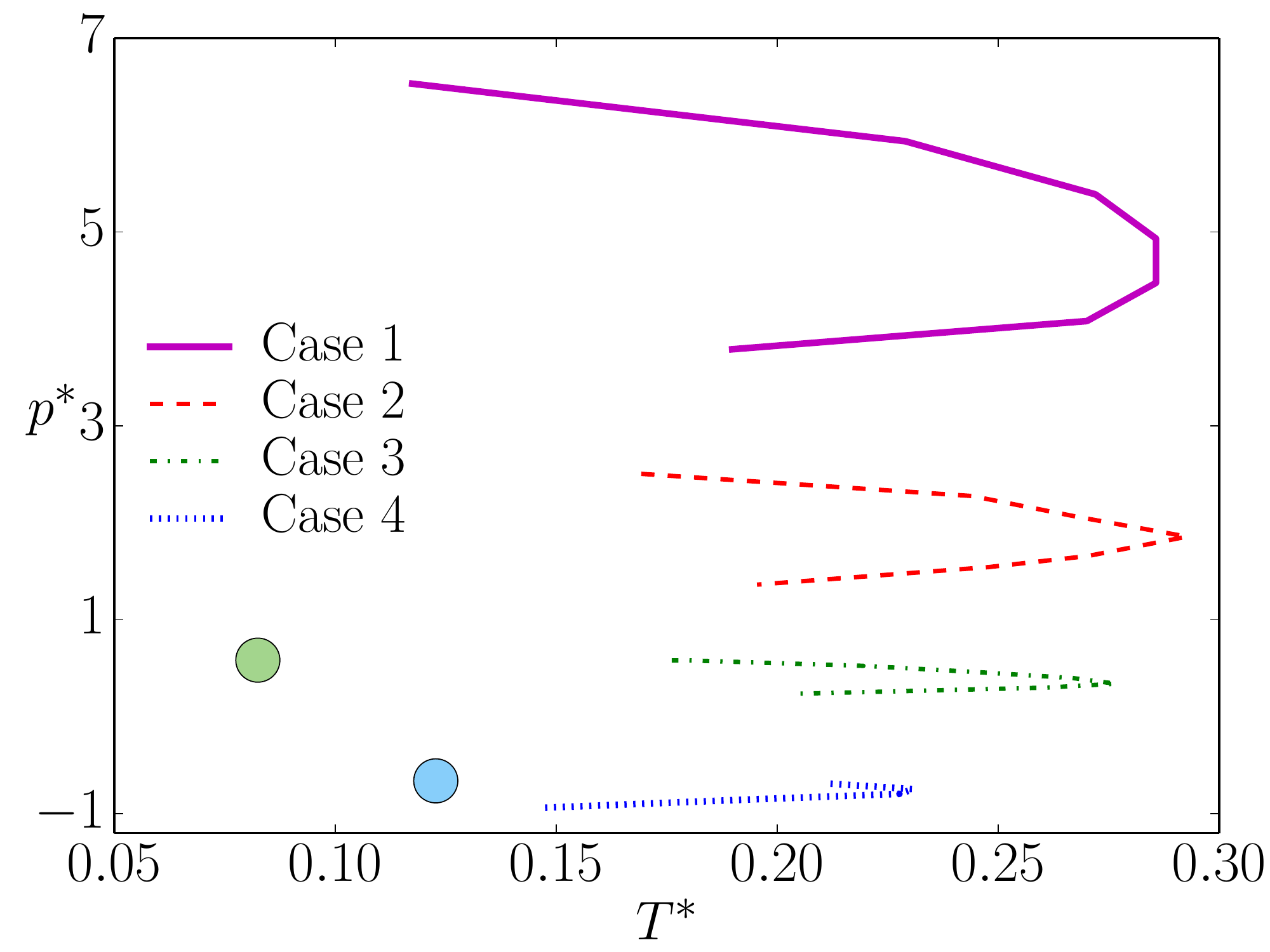}

  \caption{Pressure versus temperature phase diagram showing the TMD
    lines for all the potentials studied and critical points for the
    case 3 (green) and case 4 (blue).}
  \label{fig:TMD}
\end{figure}
%%%%%%%%%%%%%%%%%%%%%%%%%%%%%%%%%%%%%%%%%%%%%%%%%%%%%%%%%%%%%%%%%%%%%%

The figure~\ref{fig:all_pxyt} illustrates the pressure
versus temperature phase diagram for the four potentials
analyzed. As the attractive part of the
potential becomes deeper, the liquid-liquid critical
point appears. The appearance of the liquid-liquid
phase transition is 
also related to the shrink in pressure of the 
TMD line as shown by the figure~\ref{fig:TMD}. Our results supports previous
results that indicate that the presence of two
length scales with one attractive part is 
necessary but not sufficient for
the existence of two liquid phases~\cite{Fo08,Si10}. 

The condition, as suggested by  Jagla~\cite{Ja04},
for the presence of the two liquid phases is that 
the density  for fixed
temperatures has to show a non monotonic behavior with pressure.
Here we explore
if this condition also
holds for our system in which the two length scales 
are present but with continuous forces~\ref{fig:comp_lor+jagla}.

\begin{figure}[!htb]
  \includegraphics[scale=.4]{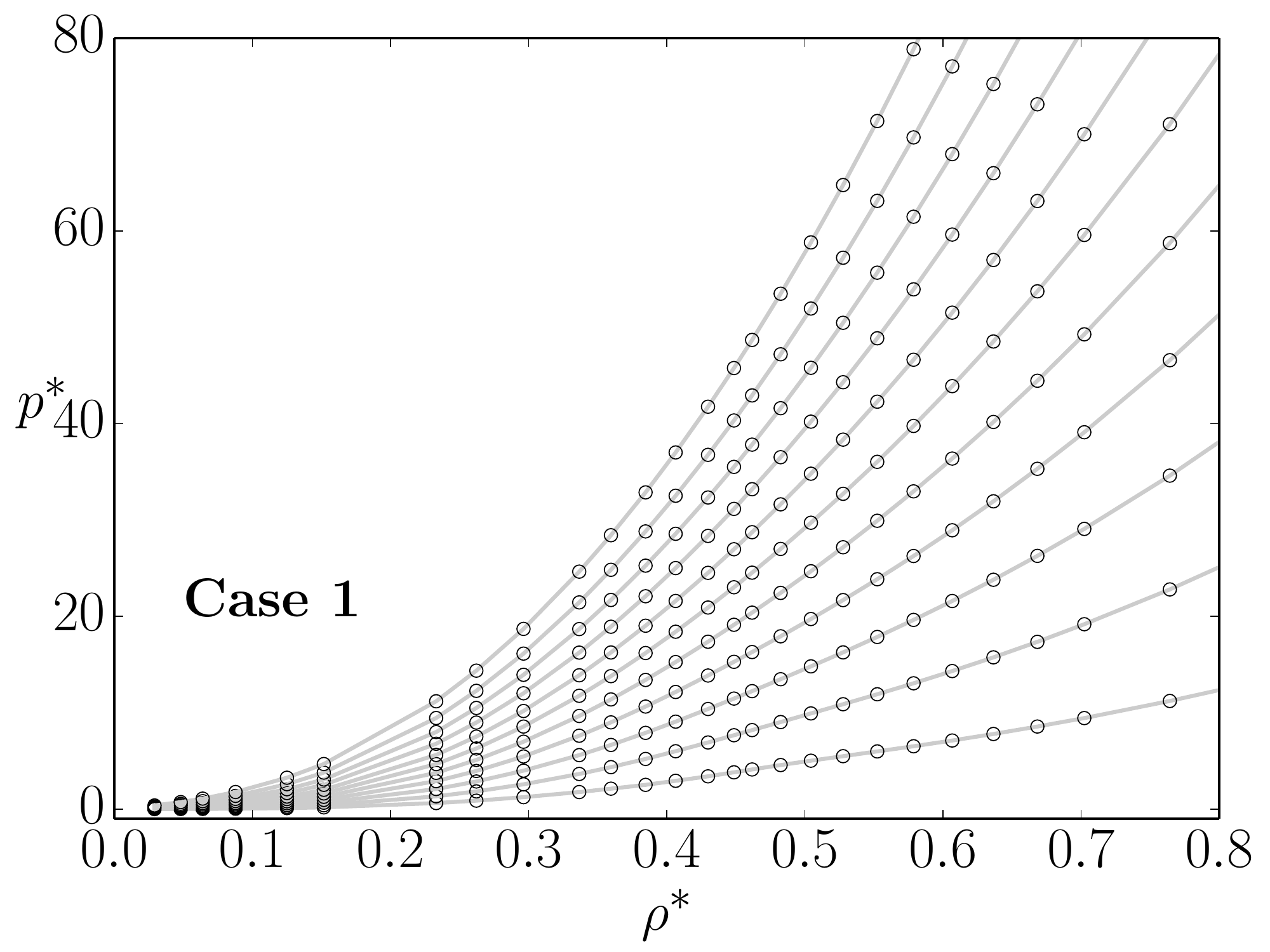}
  \includegraphics[scale=.4]{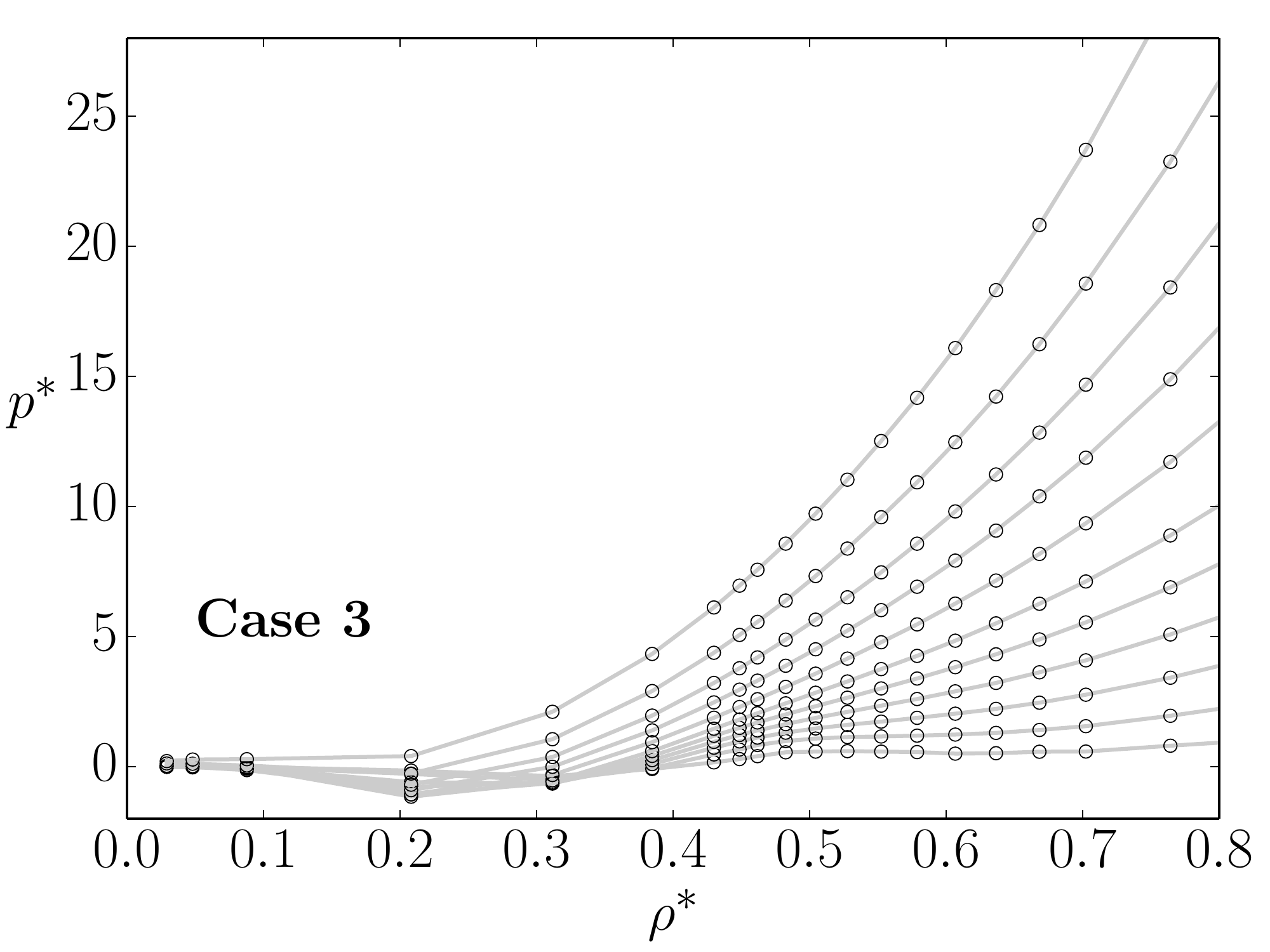}
  \caption{Pressure versus density phase diagram for the potentials
case
      1 (left) and the case 3 (right). The circles represent the 
      simulation data and the solid gray lines are guides to the 
      eyes. On the insets, we show the region where the isotherms 
      cross. The range of the temperatures shown on the insets are 
      $0.1<T^*<0.4$ and $0.15<T^*<0.5$ for cases 1 and 3,
      respectively.}
  \label{fig:all_pxyrho}
\end{figure}

Figure~\ref{fig:all_pxyrho} illustrates the pressure versus
density isotherms.
For clarity the pressures were shifted 
as $p^*=n\times p^*$, where $n=1,2,\ldots$ are used
for increasing temperatures.
In our system, as in 
the Jagla's potential the  increase of the attractive
part of the pair potential contributes to negative
pressure until a critical value.
In this region  the  density versus 
pressure at constant temperature becomes
reentrant, resulting in a first-order phase transition and a second
critical point.
In the case 1 where no liquid-liquid 
critical point is present the density 
increases with the pressure monotonically
while in the case 3 in which the critical point appears
in a range of temperatures the pressure  becomes
reentrant.

%%%%%%%%%%%%%%%%%%%%%%%%%%%%%%%%%%%%%%%%%%%%%%%%%%%%%%%%%%%%%%%
%%%%%%%%%%%%%%%%%%%%%%%%%%%%%%%%%%%%%%%%%%%%%%%%%%%%%%%%%%%%%%%

Although the anomaly in density is present in all the 
potentials analyzed, as the potential becomes
more attractive, the region in pressure occupied 
by the TMD decreases. This result is a consequence
of the link between the TMD curvature and the 
presence of phase separation.

Another consequence of
the link between criticality and the anomalous behavior
is the region in pressure occupied by the dynamic 
and structural anomalies illustrated as the 
dashed and dotted lines respectively in the figure~\ref{fig:all_pxyt}
shrinks in pressure as the potential becomes more attractive.
These thermodynamic, dynamic and
structural anomalies are related by the radial distribution
function as follows.
The radial distribution function $g(r)$ is a measure of a probability
of finding a pair of particles at a given distance $r$, which can be
evaluated by equation~(\ref{eq:g_of_r}), and its
behavior is a key ingredient for the presence of 
the anomalies.

%%%%%%%%%%%%%%%%%%%%%%%%%%%%%%%%%%%%%%%%%%%%
\begin{figure}[!htb]
  \includegraphics[scale=.4]{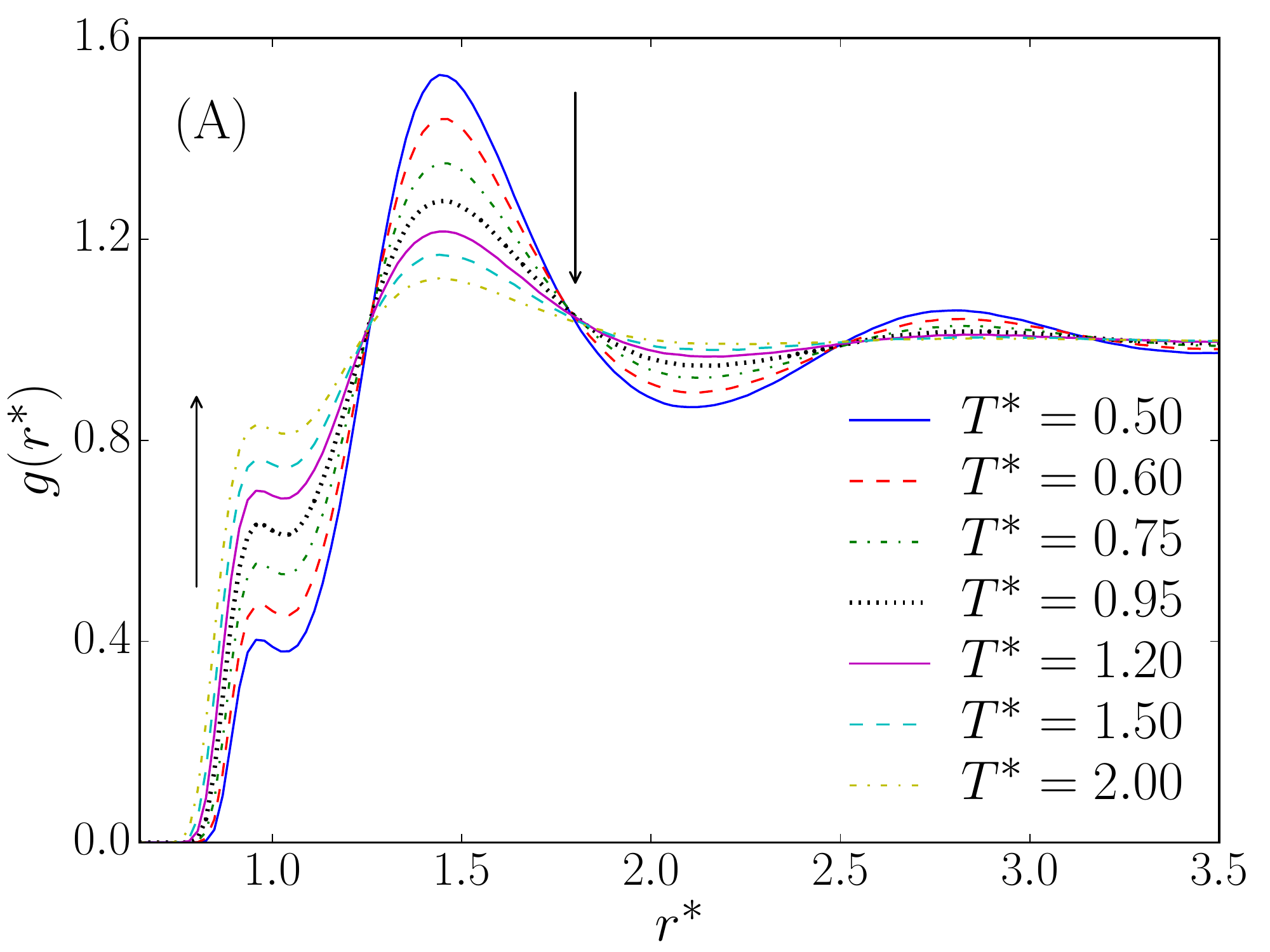}
  \includegraphics[scale=.4]{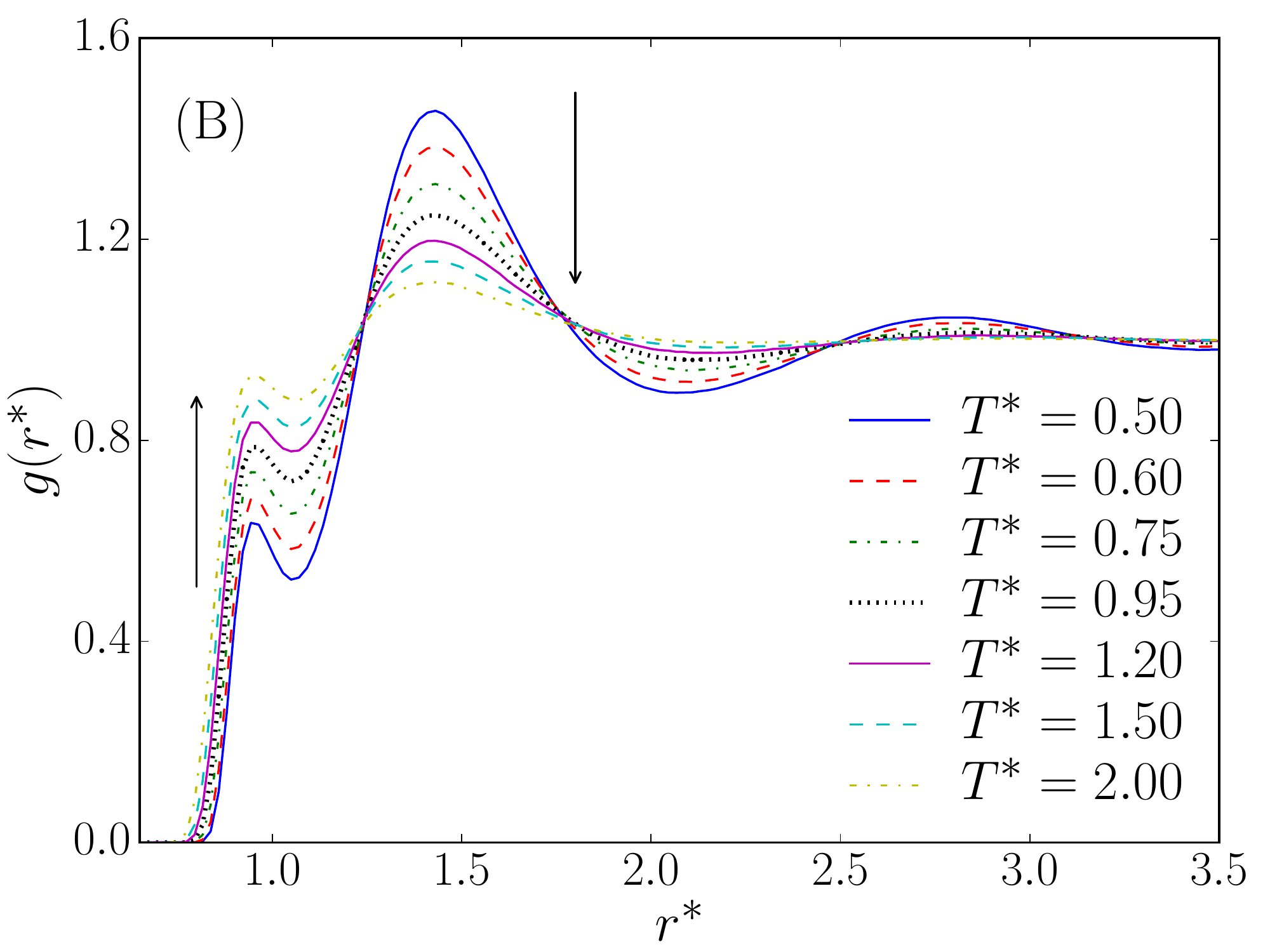}
     \caption{Radial distribution function as a function of radial
       distance, for the (A) Case 3 at $\rho^*=0.380$ and (B) Case 4 at
       $\rho^*=0.430$. The arrows indicate the inversion of the 
highest value of $g(r^*)$
       for each isotherm analyzed.}
    \label{fig:comp_gr}
\end{figure}
%%%%%%%%%%%%%%%%%%%%%%%%%%%%%%%%%%%%%%%%%%%%

The figure~\ref{fig:comp_gr} illustrates
the $g(r)$ as a function of the distance for
different temperatures for
the potentials case less ideal-gas-like with the increase of the 
density.  As the temperature is
increased the mean distance between
particles decreases and, as a consequence, the first peak of $g(r)$ increases
while the second peak decreases. Particles changing from the 
one length scale to the other is the characteristic of 
the density anomaly.~\cite{Ol09,Ba09,Si10,Ba11}.

The unusual behavior of the $g(r)$ also reflects in the structure.
The translational order parameter
$t^*$, as defined in the equation (\ref{eq:trans_ord_param})
measures how structured is the system. For the ideal gas, for example,
we have $g(\xi)=1$, $\forall \xi$, and thus $t=0$. 
%%%%%%%%%%%%%%%%%%%%%%%

%%%%%%%%%%%%%%%%%%%%%%%Meanwhile, for
crystallized systems, where the particles have a well defined
structure, we have $g(\xi) \neq 1$ and then $t \neq 0$. Therefore, for
normal liquids, $t$ increases with the increasing of density, since an
increasing in density induces structuration in system.
%%%%%%%%%%%%%%%%%%%%%%%%%%%%%%%%%%%%%%%%%%%
\begin{figure}[!htb]
    \includegraphics[scale=.35]{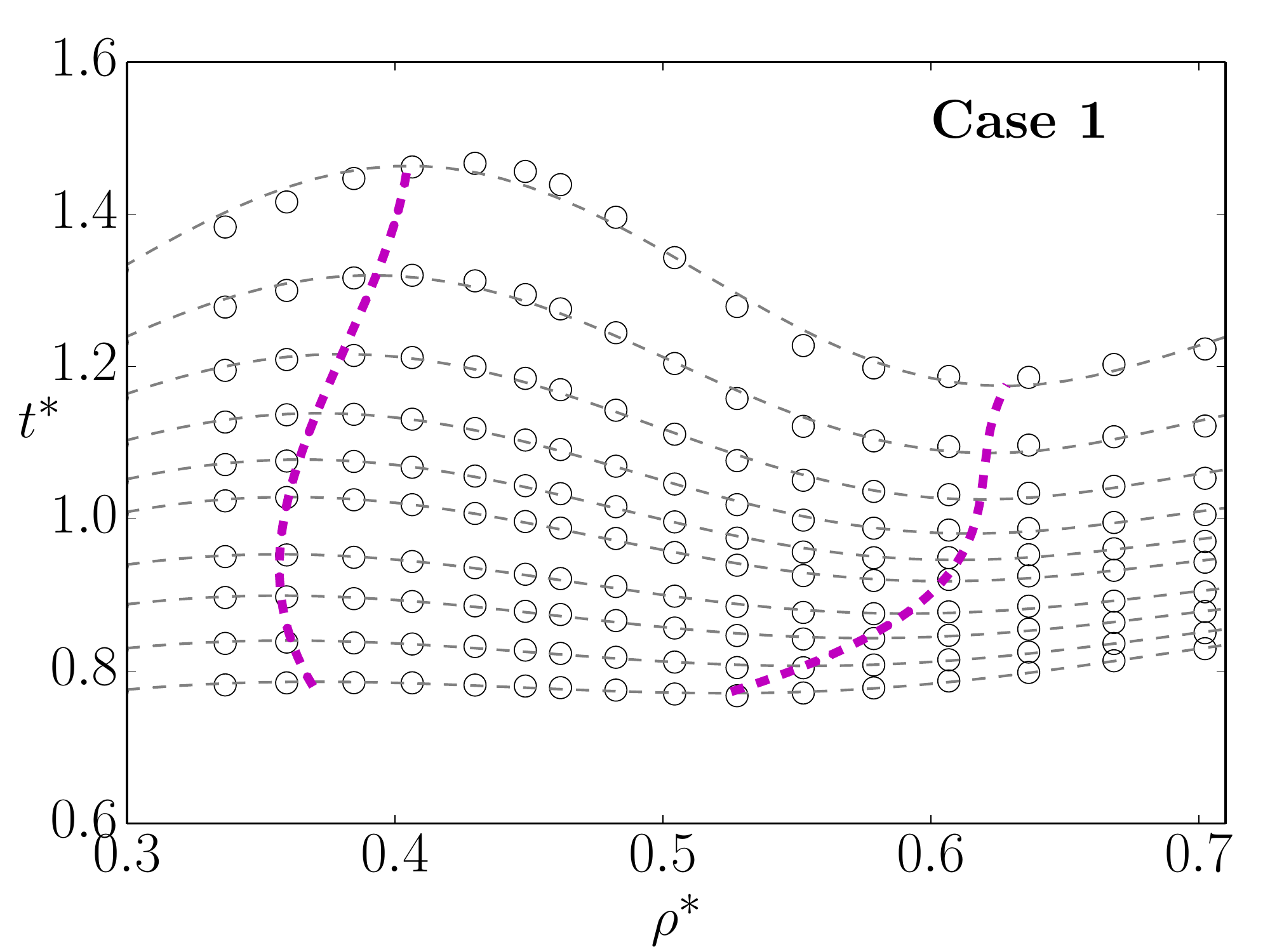}
    \includegraphics[scale=.35]{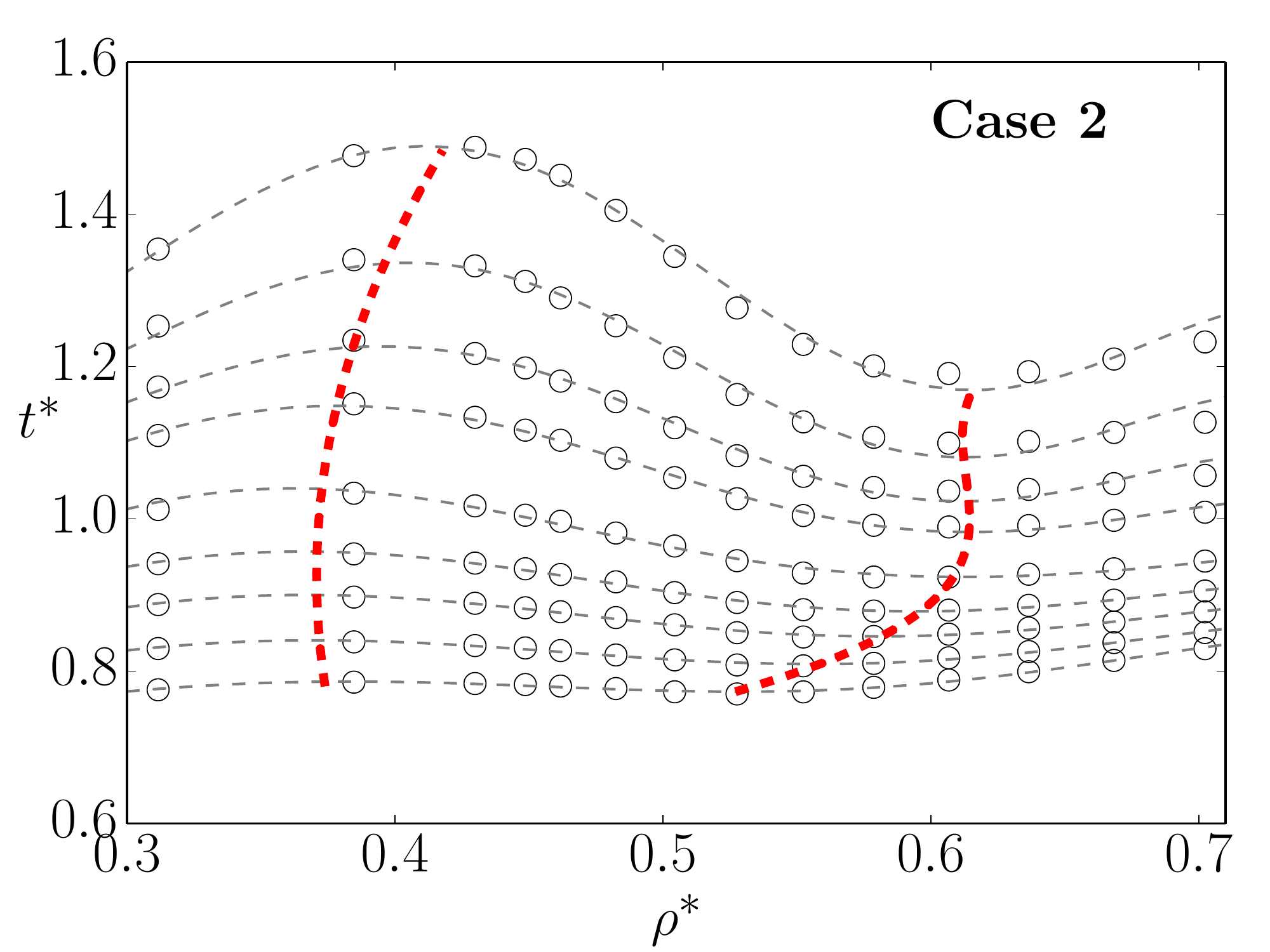}

    \includegraphics[scale=.35]{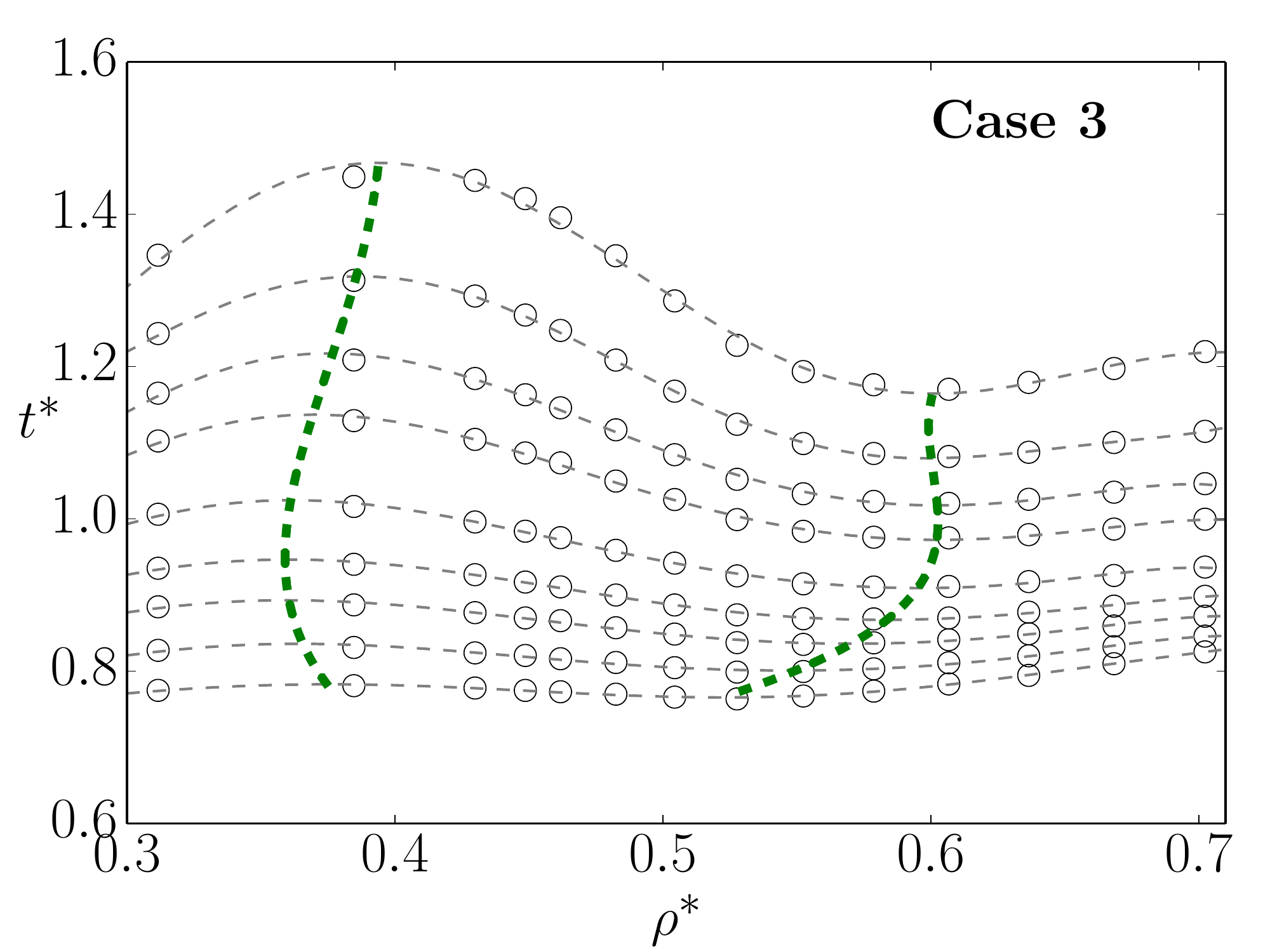}
    \includegraphics[scale=.35]{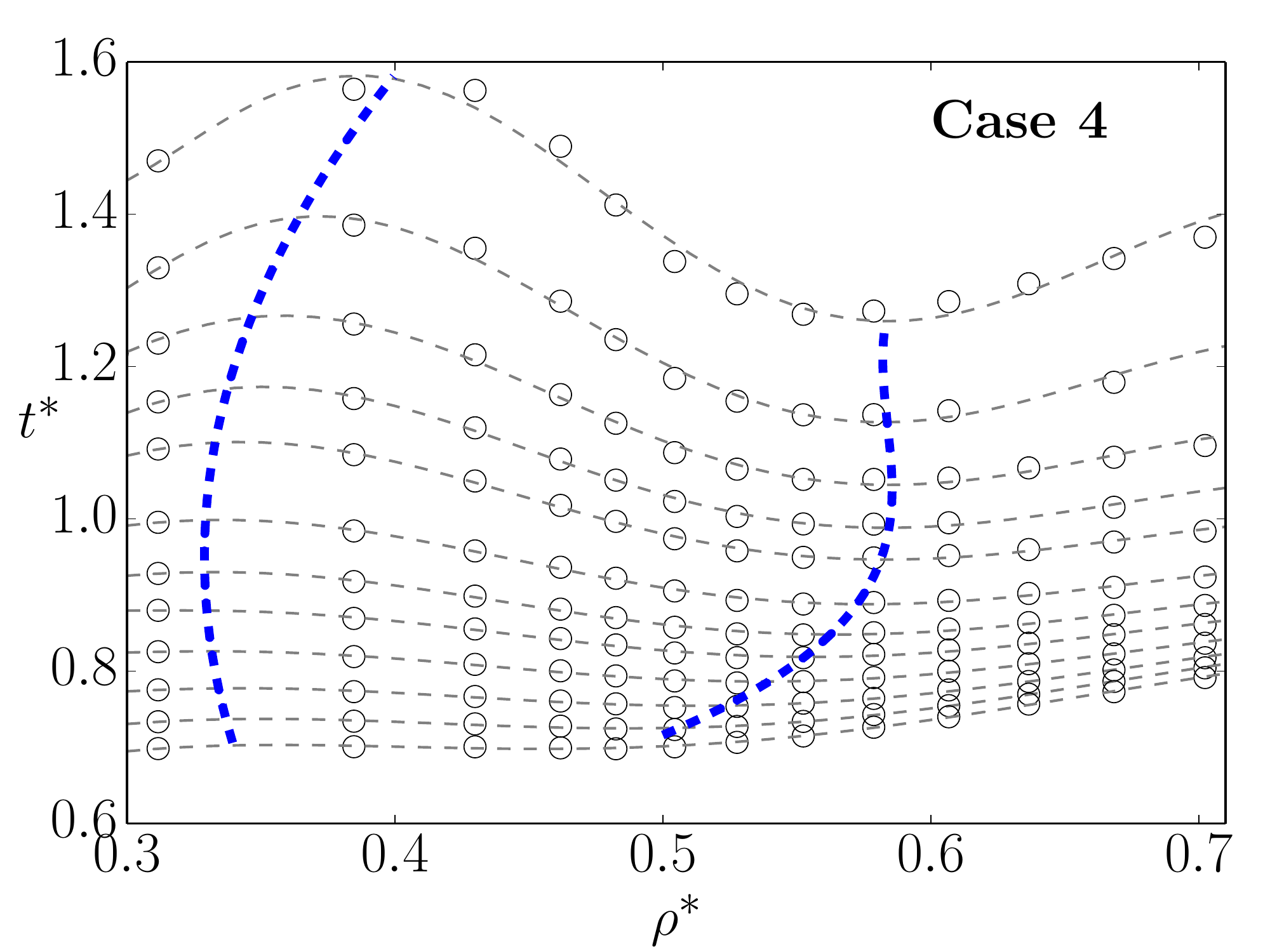}

    \caption{Translational order parameter versus
density   for  $T_{1}^*=0.100,
      0.150,\ldots,0.300,0.400,\ldots,0.950,1.20,1.50$, $T_{2}^*=0.15,
      0.20,0.25,\ldots,0.95$ , $T_{3}^*=0.15,0.20, 0.25,\dots,1.50$ ,
      $T_{4}^*=0.15,0.20,0.25,\ldots, 0.95$ for
the four potentials. The solid gray lines are
      polynomial fit and circles are points obtained by simulation.
      Dashed lines denotes the limit of the anomalous region.}
    \label{fig:all_t}
%  \label{fig:all_t}
\end{figure}
%%%%%%%%%%%%%%%%%%%%%%%%%%%%%%%%%%%%%%%%%%%

The figure~\ref{fig:all_t} shows the translational order parameter as a
function of the reduced density for fixed temperatures for all the
cases studied. There is a region where in densities 
in which the parameter
$t^*$ decreases as the density increases, what
is the signature of the anomaly.  The increase of the
attractive part in the potential hinders the movement between the
two scales scales, reducing the manifestation of the anomaly in $t$. Similar
results were obtained by Barraz {\it et al.}~\cite{Ba09} for an
isotropic water-like model.

\begin{figure}[!htb]
    \includegraphics[scale=.35]{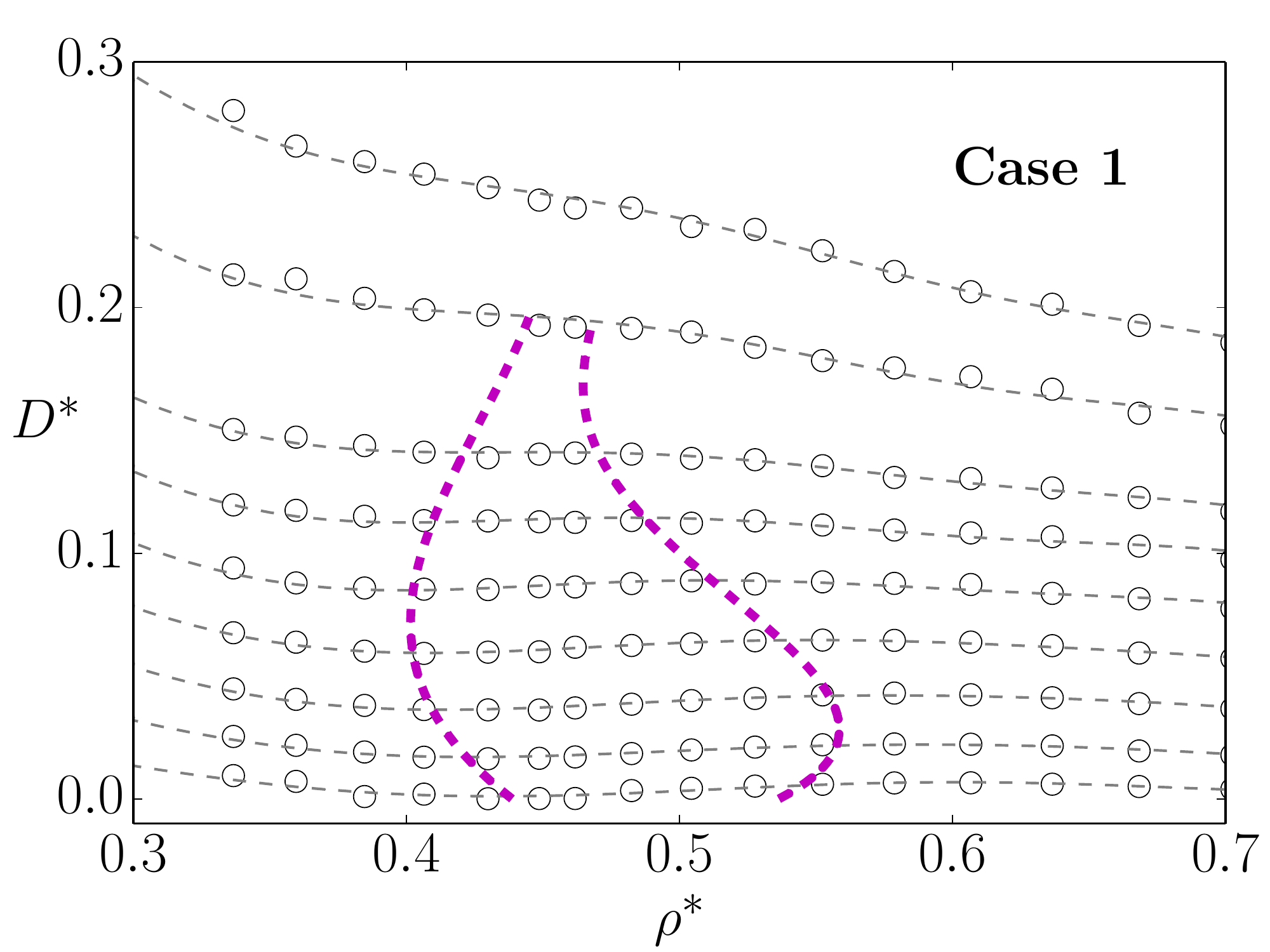}
    \includegraphics[scale=.35]{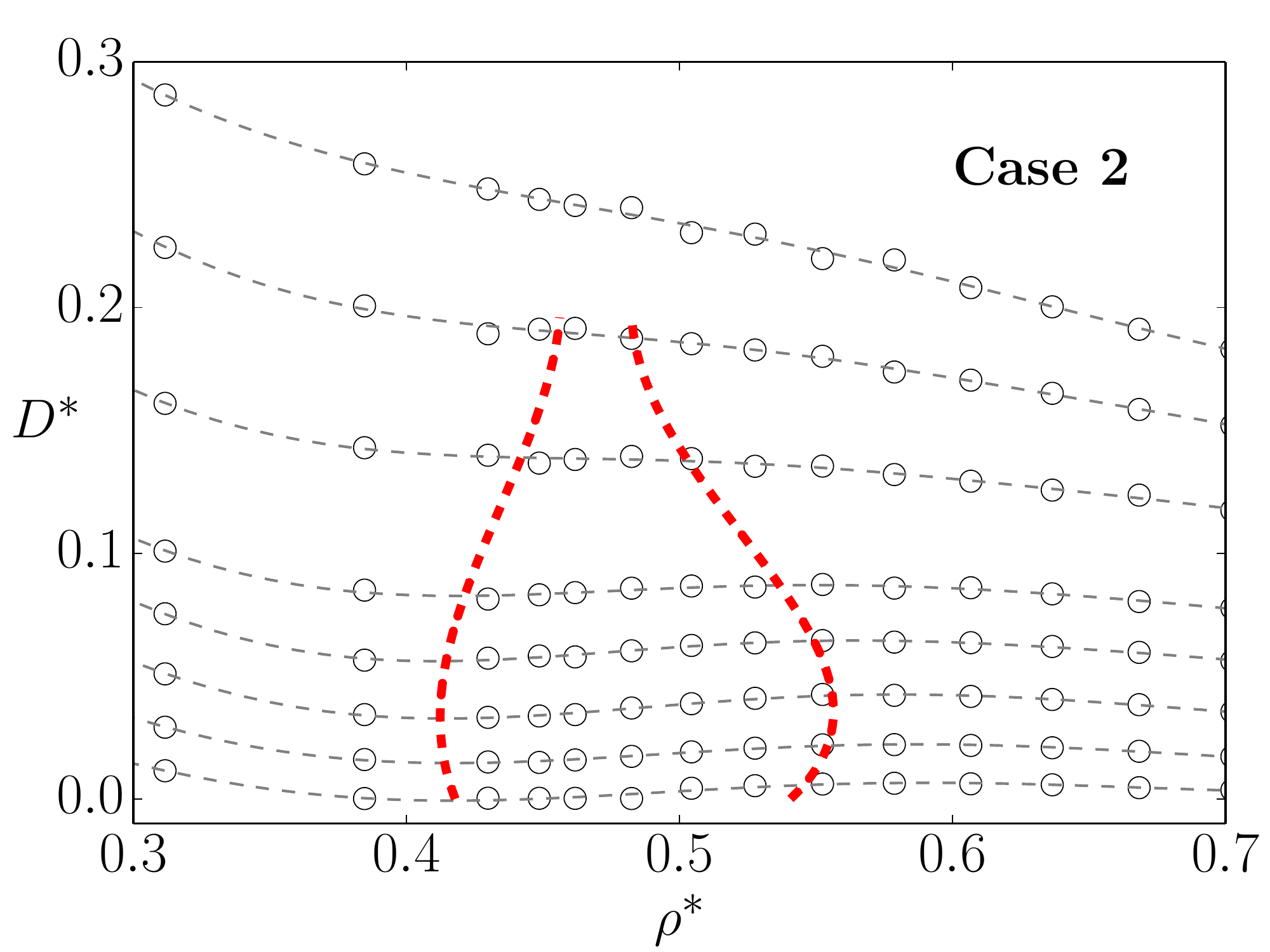}

    \includegraphics[scale=.35]{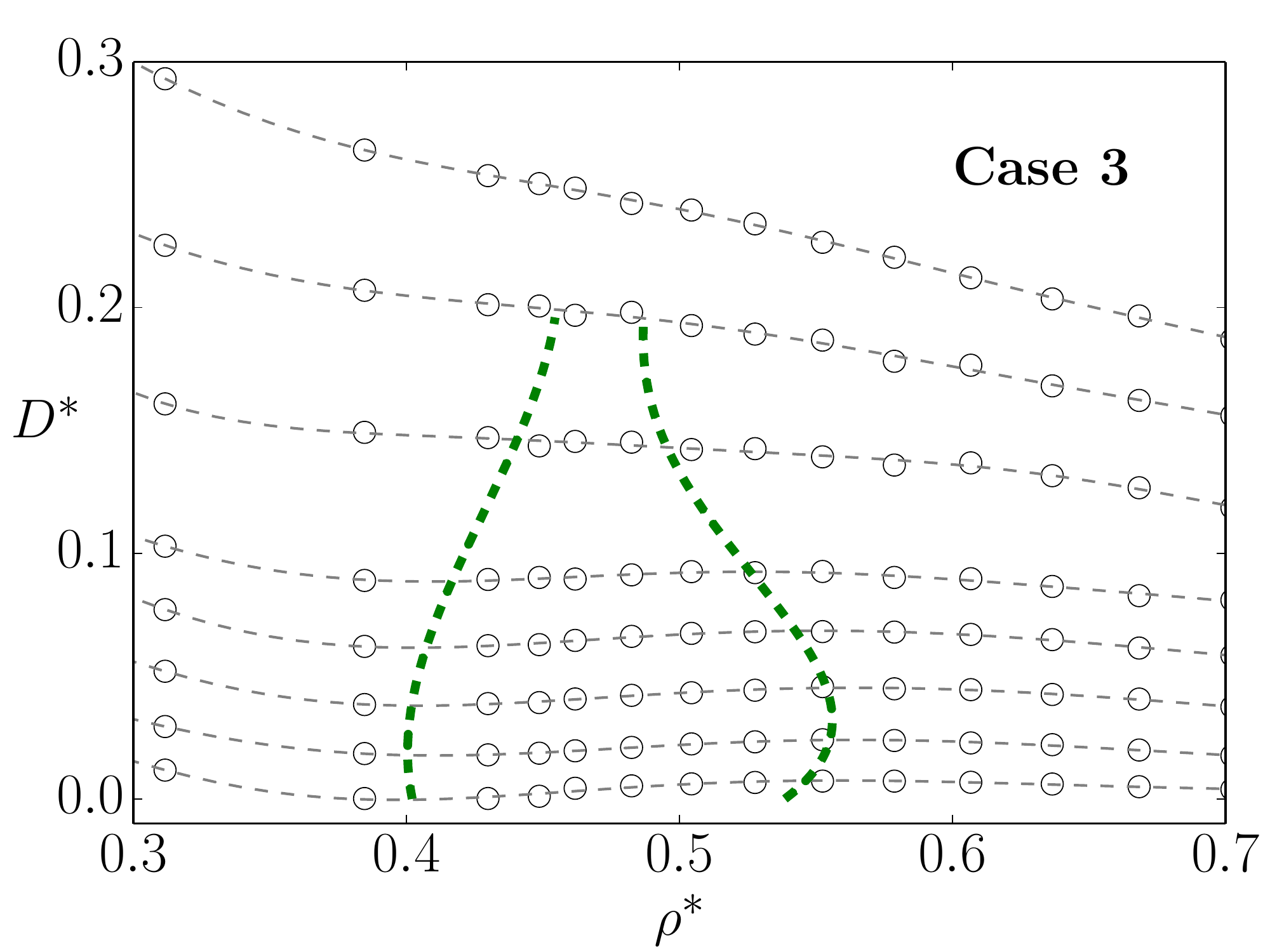}
    \includegraphics[scale=.35]{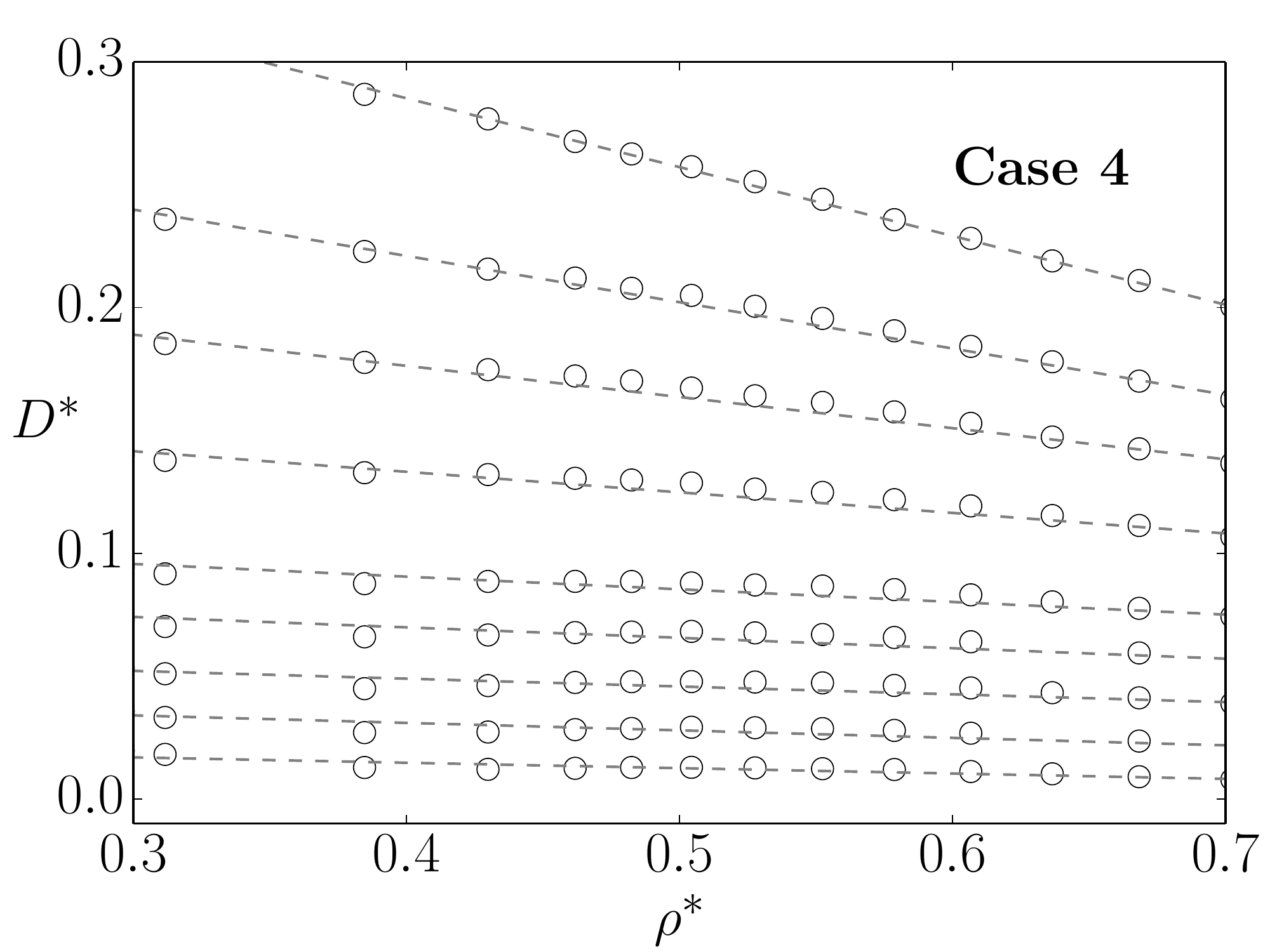}

    \caption{Diffusion coefficient as function of density 
for  $T_{1}^*=0.20,\ldots,2.50$,
      $T_{2}^*=0.05,\ldots,2.500$ ,$T_{3}^*=0.5,\ldots,2.50$,
      $T_{4}^*=0.05,\ldots,2.50$ and $T^*_{5}=0.1,\ldots,1.05$
for all the potentials studied. The
      solid gray lines are polynomial fits and circles are points
      obtained by simulations. Dashed lines denote the region of the
      anomalous behavior in $D^*$.}
    \label{fig:all_diff}
\end{figure}

The diffusion is computed from equation~\ref{difusao_lateral}.
For normal systems the diffusion decreases with the increase
of the density. 
Figure~\ref{fig:all_diff} shows the behavior of the translational
diffusion coefficient $D^*$, as a function of the reduced density
$\rho^*$, at constant temperature. For the cases 1, 2 and 3
the systems show a region in which the diffusion increases
with the increase of the density what characterizes 
the diffusion anomaly. As the attractive part of the 
potential becomes larger, the value of $D$ in which the 
anomalous behavior is observed decreases and the 
region in pressures for the anomalous behavior also shrinks.
The mobility of the particles are strongly affected by
the depth of the potential.

%%%%%%%%%%%%%%%%%%%%%%%%%%%%%%%%%%%%%%%%%%%%%%%%
%%%%%%%%%%%%%%%%%%%%%%%%%%%%%%%%%%%%%%%%%%%%%%%%

Here we study a excess entropy defined by equation~(\ref{eq:s2}). The
excess entropy measures the decreasing of the entropy of the real
liquid, when compared to an ideal gas at the same temperature and density,
due to structural correlations. If a system has no anomaly, there
is no preference of particles to assume a specific coordination
shell~\cite{Ol10b}. However, we observe that for our model the
particles move from the second coordination shell to the first coordination
one. The structural correlation between the particles can be captured
by the excess entropy.

\begin{figure}[!htb]
  \includegraphics[scale=.35]{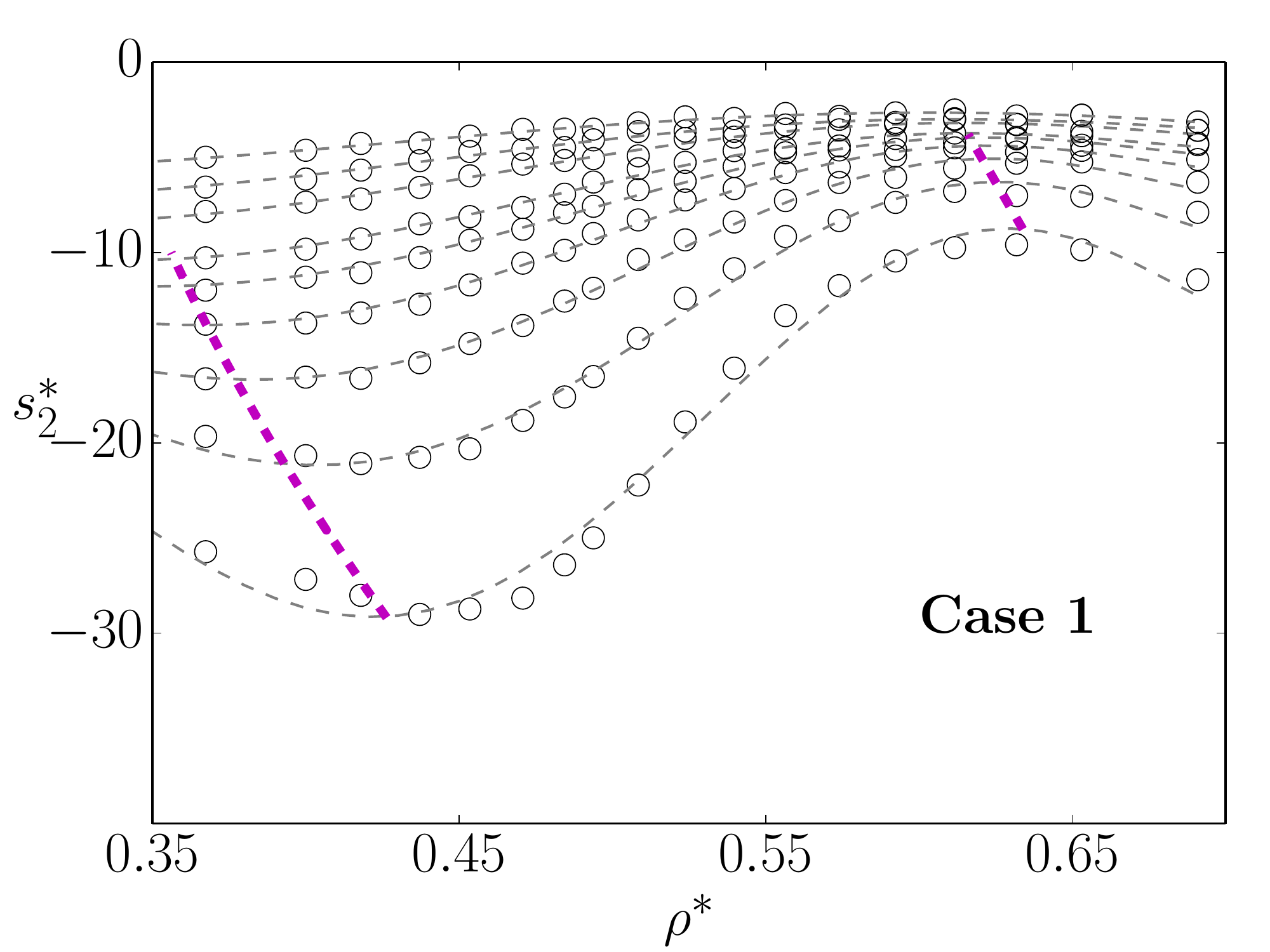}
  \includegraphics[scale=.35]{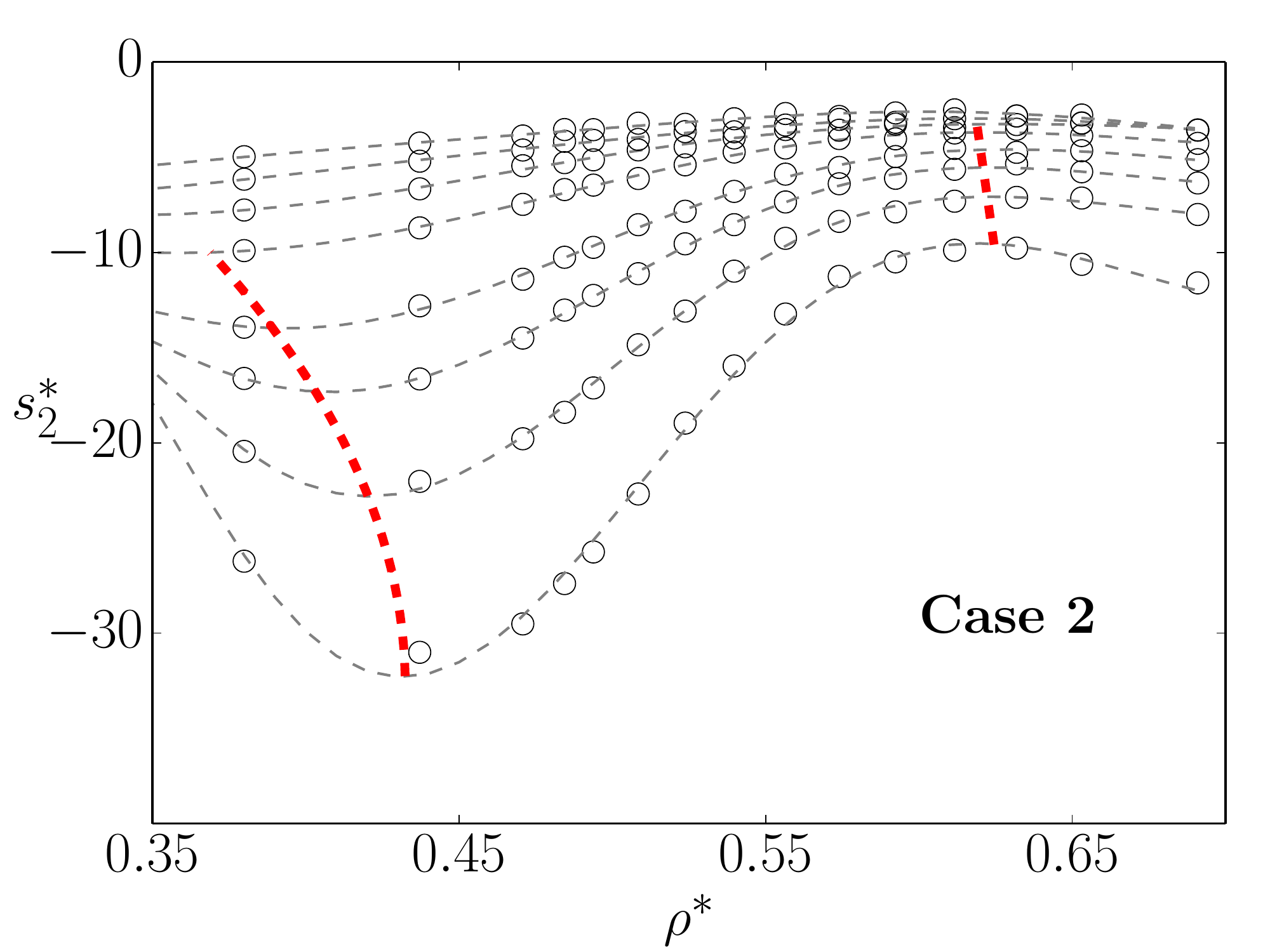}

  \includegraphics[scale=.35]{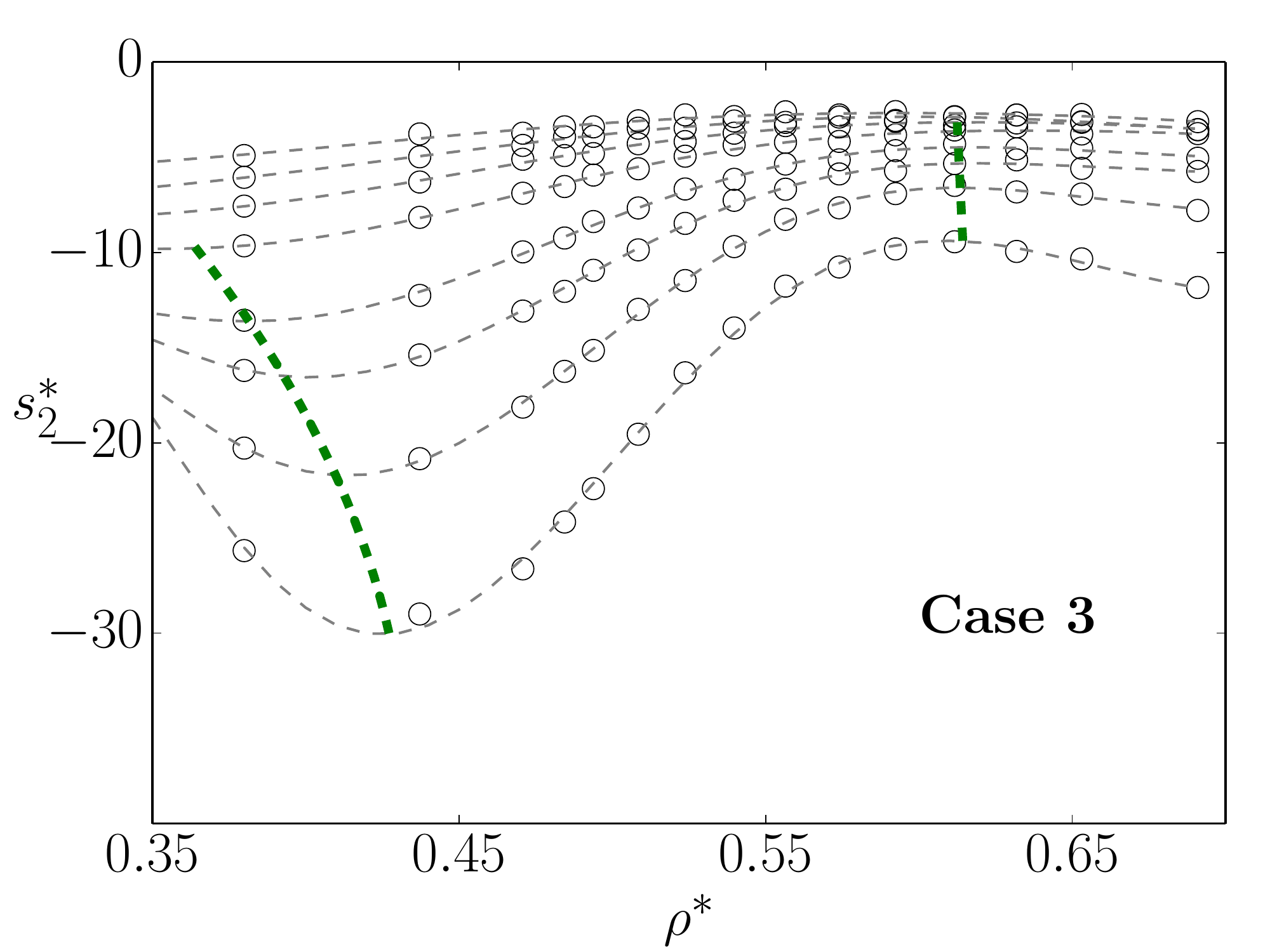}
  \includegraphics[scale=.35]{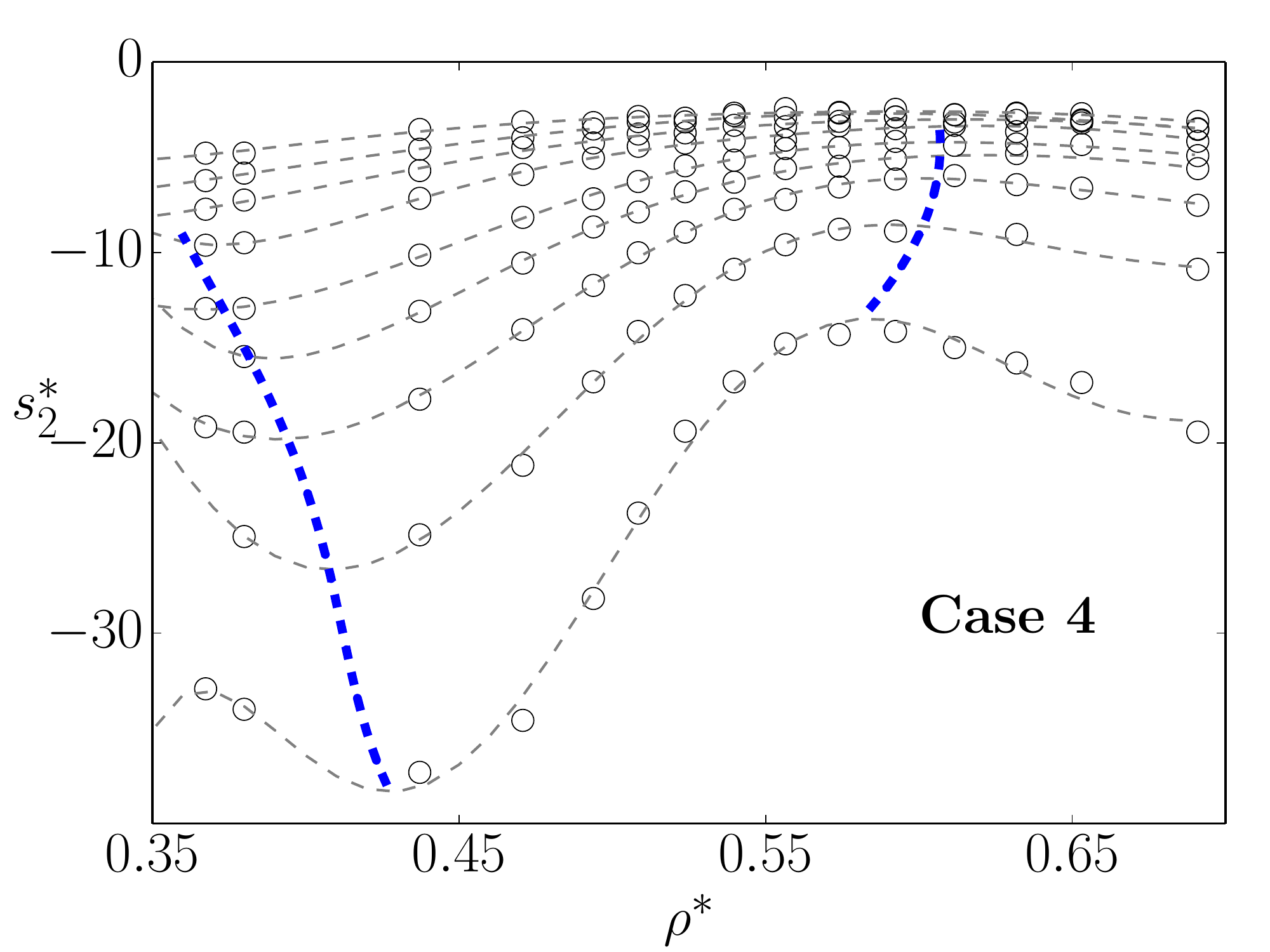}
  \caption{Excess entropy versus density for $T_{1}^*=0.20,\ldots,2.50$,
      $T_{2}^*=0.05,\ldots,2.500$ ,$T_{3}^*=0.5,\ldots,2.50$,
      $T_{4}^*=0.05,\ldots,2.50$ and $T^*_{5}=0.1,\ldots,1.05$. The gray 
solid lines
    are polynomial fit and circles are simulational points. Dashed 
lines comprise the anomalous region. }
  \label{fig:all_s2}
\end{figure}

Related to the 
hierarchy of anomalies 
the excess entropy is 
computed from equation~\ref{eq:s2}.
For normal liquids the excess of entropy decreases
with the increase of the density since the system
becomes more structured with the increase of density.
Figure~\ref{fig:all_s2} shows that for
our potentials there is a region in densities
in which the excess entropy increases  with the
increase of density what characterizes the region 
of densities in which the system has anomalous behavior.
As in the density, diffusion and translational anomalous 
behavior, this region shrinks in pressure range as 
the system becomes more attractive.

%%%%%%%%%%%%%%%%%%%%%%%%%%%%%%%%%%%%%%%%%%%%%%%%
%%%%%%%%%%%%%%%%%%%%%%%%%%%%%%%%%%%%%%%%%%%%%%%%

The use of the NVT molecular dynamic described above
is very useful for understanding the anomalous behavior 
and for locating the critical point. However this 
method is not manageable for obtaining the coexistence 
line and the universality class of the transition. 

\begin{figure}[!htb]
  \includegraphics[scale=0.5]{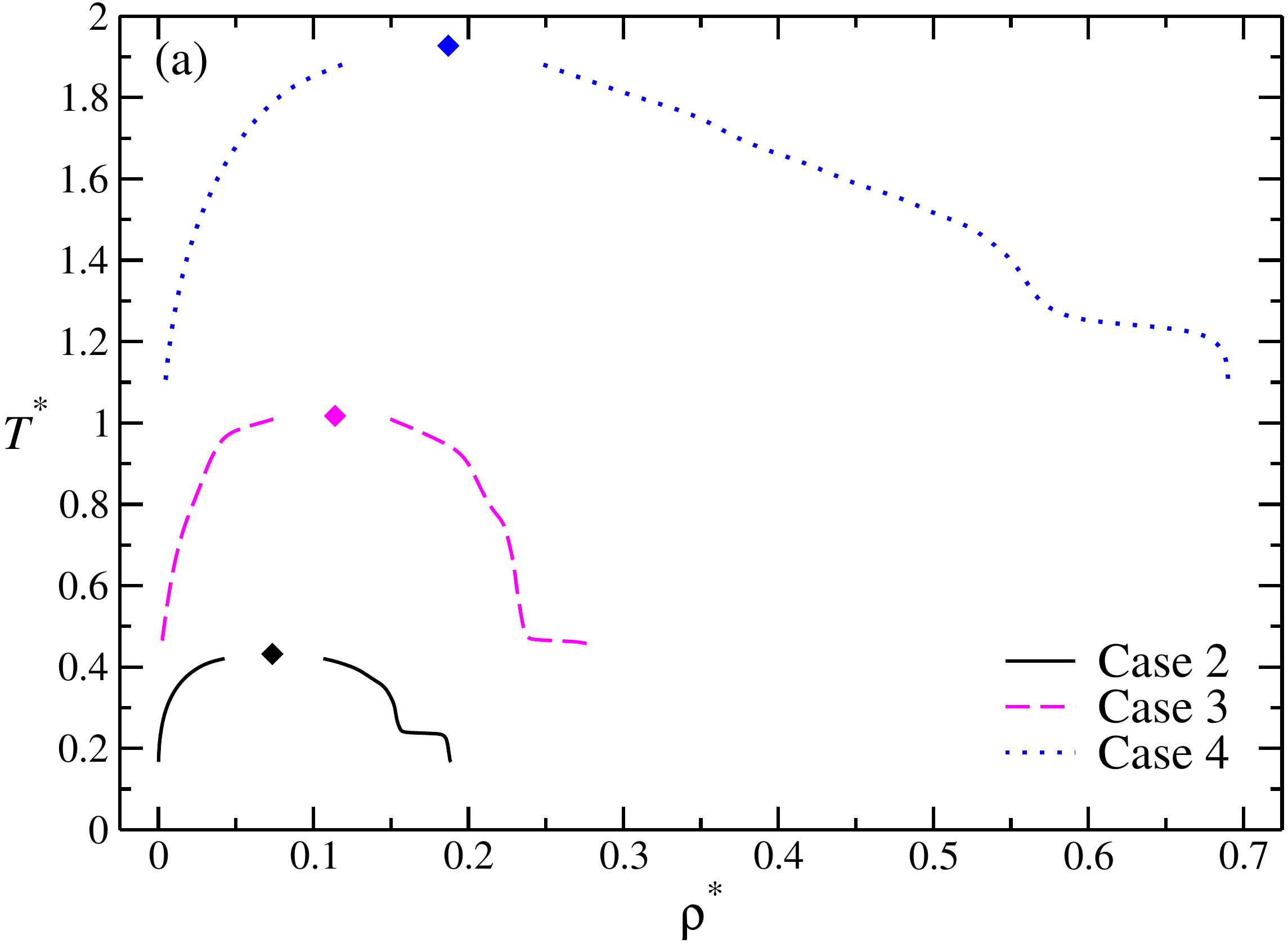}
  \caption{Density versus temperature phase diagram  for
the liquid gas coexistence for the three potentials
with an attractive part.}
\label{fig:txrho_phase}
\end{figure}

In order to understand the 
nature of the phases produced
by the effective potentials studied 
in the work, Grand Canonical Monte Carlo analysis
was employed.
First, the liquid-gas phase transition was analyzed.
 Using the estimates of the critical region and critical
histograms obtained from simulations, the low and the high density
histograms were combined  to obtain the coexistence region, as shown in
figure~\ref{fig:txrho_phase}.

%%%%%%%%%%%%%%%%%%%%%%%%%%%%%%%%%%%%%%%%%%%%%%%%%%%%%%%%%%%%%%%%
%%%%%%%%%%%%%%%%%%%%%%%%%%%%%%%%%%%%%%%%%%%%%%%%%%%%%%%%%%%%%%%%

%%%%%%%%%%%%%%%%%%%%%%%%%%%%%%%%%%%%%%%%%%%%%%%%%%%%%%%%%%%%%%%%
\begin{figure}[!htb]
  \includegraphics[scale=0.5]{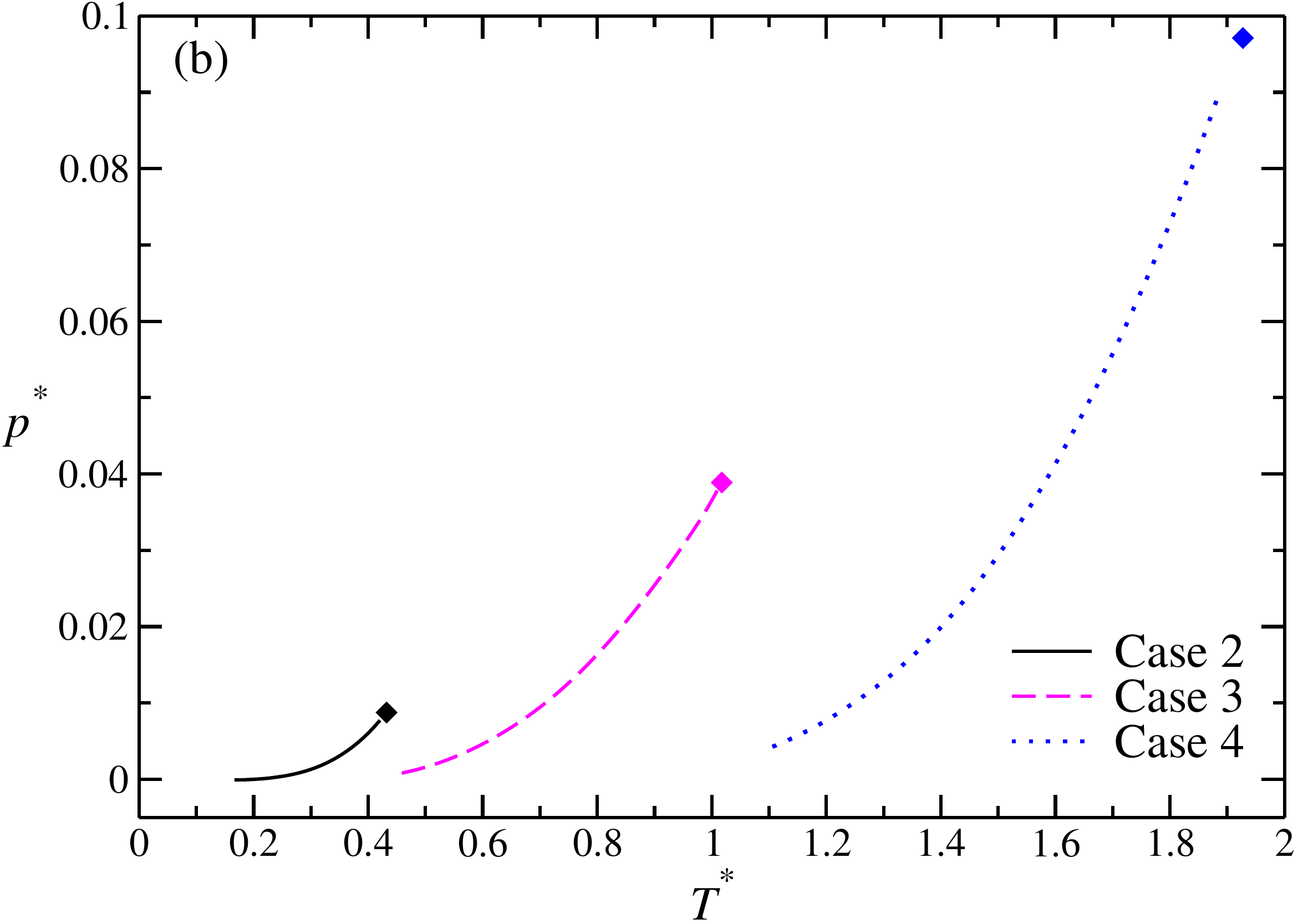}
  \caption{Pressure versus temperature phase diagram  for
the liquid gas coexistence for the three potentials
with an attractive part.}
\label{fig:pxt_phase}
\end{figure}
%%%%%%%%%%%%%%%%%%%%%%%%%%%%%%%%%%%%%%%%%%%%%%%%%%%%%%%%%%%%%%%%

The density versus temperature phase diagram, figure~\ref{fig:txrho_phase},
shows  an increase of the critical temperature and
density when the attractive scale becomes deeper. 
Then using the method described 
in the section~\ref{sec:simulation} the 
pressure of the liquid-gas critical point
was obtained.  The figure~\ref{fig:pxt_phase} illustrates 
the pressure versus temperature phase diagram
of the liquid-gas coexistence and it shows that
as the attractive part of the potential becomes deeper, the 
temperature and the pressure of the liquid-gas critical point increases what
is natural since more temperature is required to form the fluid
phase.  A similar 
result was found for a spherical
potential with two length scales by MD simulations~\cite{Si10}.

%%%%%%%%%%%%%%%%%%%%%%%%%
\begin{figure}[!htb]
  \centering
  \includegraphics[scale=.4]{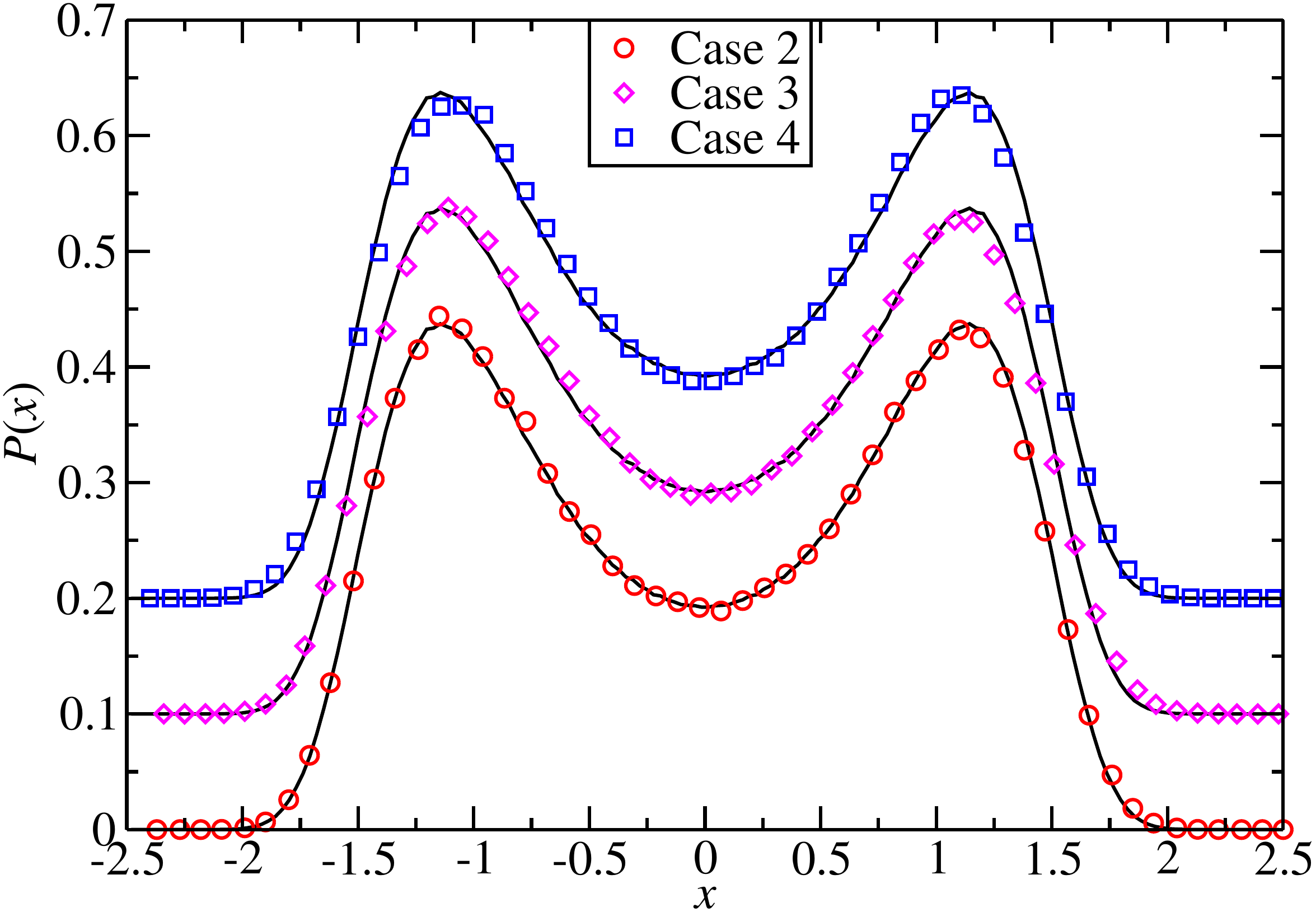}
  \caption{Comparison between the probability $P(x)$ for the
    the case 2 (circles), 3 (diamond) and 4 (squares). The full line
    represents the universal curve for the Ising 3D
    universality class for $L^{\ast}=18$.}
  \label{fig:crit_allB}
\end{figure}
%%%%%%%%%%%%%%%%%%%%%%%%%

%%%%%%%%%%%%%%%%%%%%%%%%%%%%%%%%%%%%%%%%
\begin{figure}[!htb]
\includegraphics[scale=0.35]{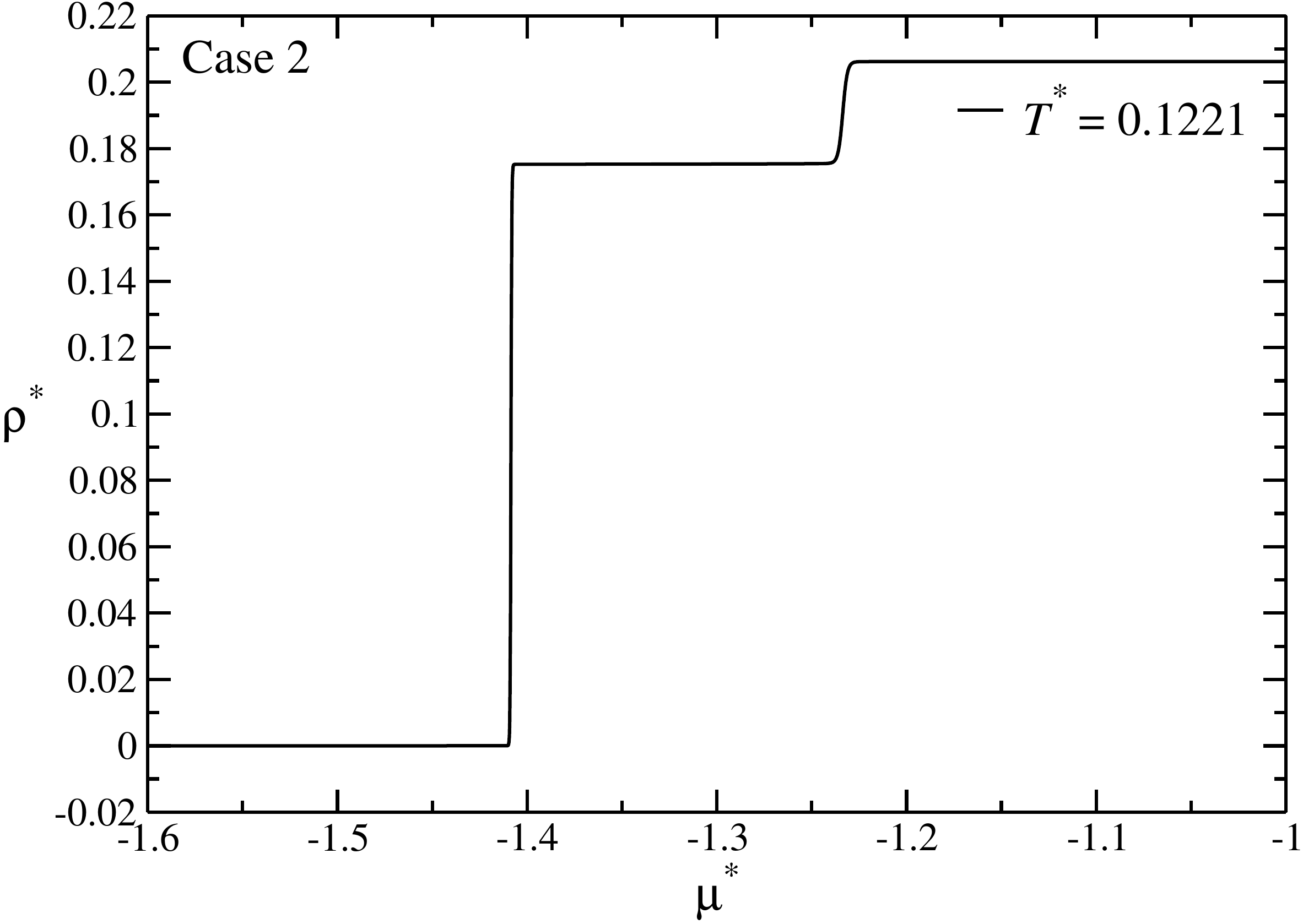}
\includegraphics[scale=0.35]{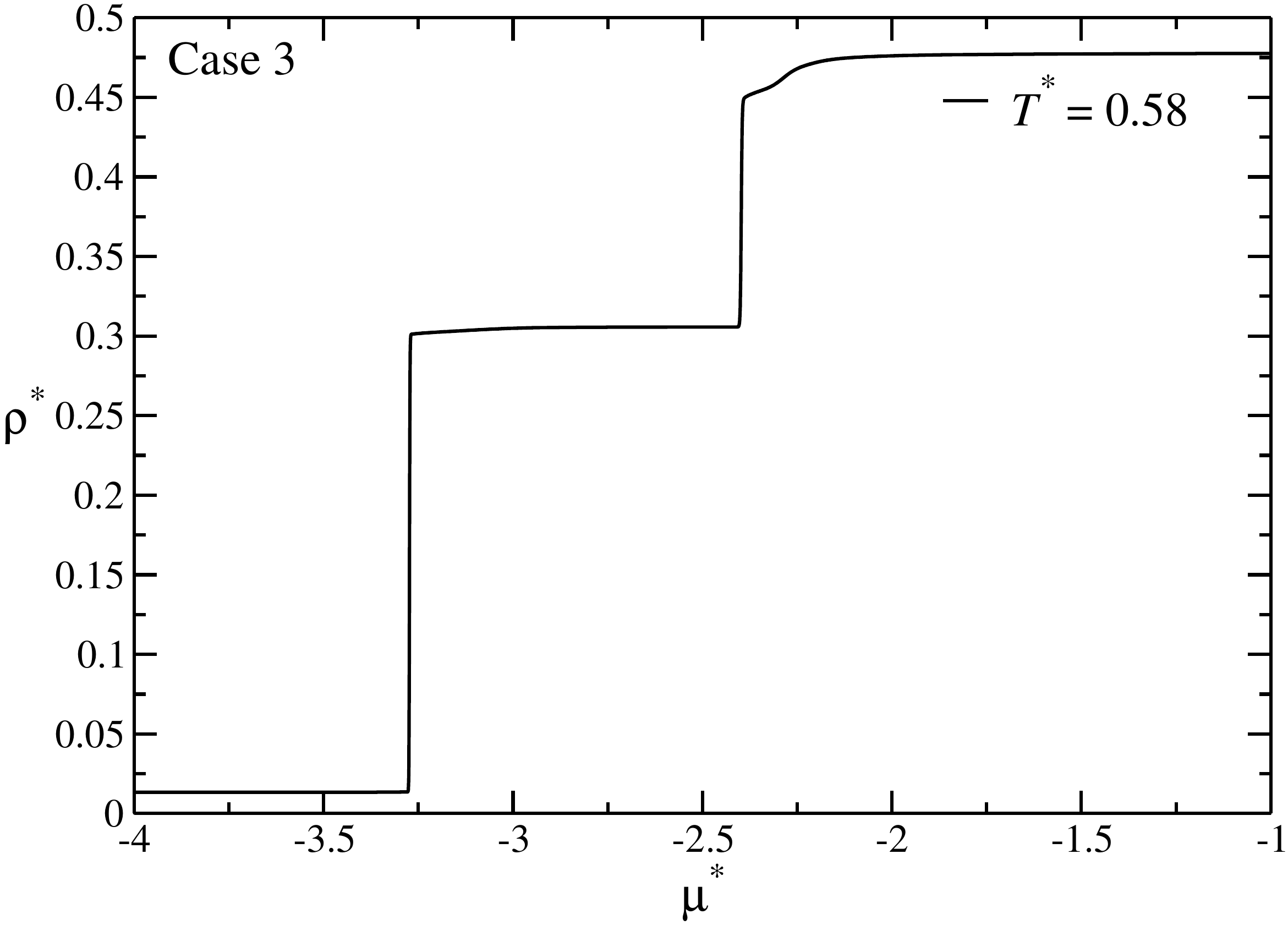}
\includegraphics[scale=0.35]{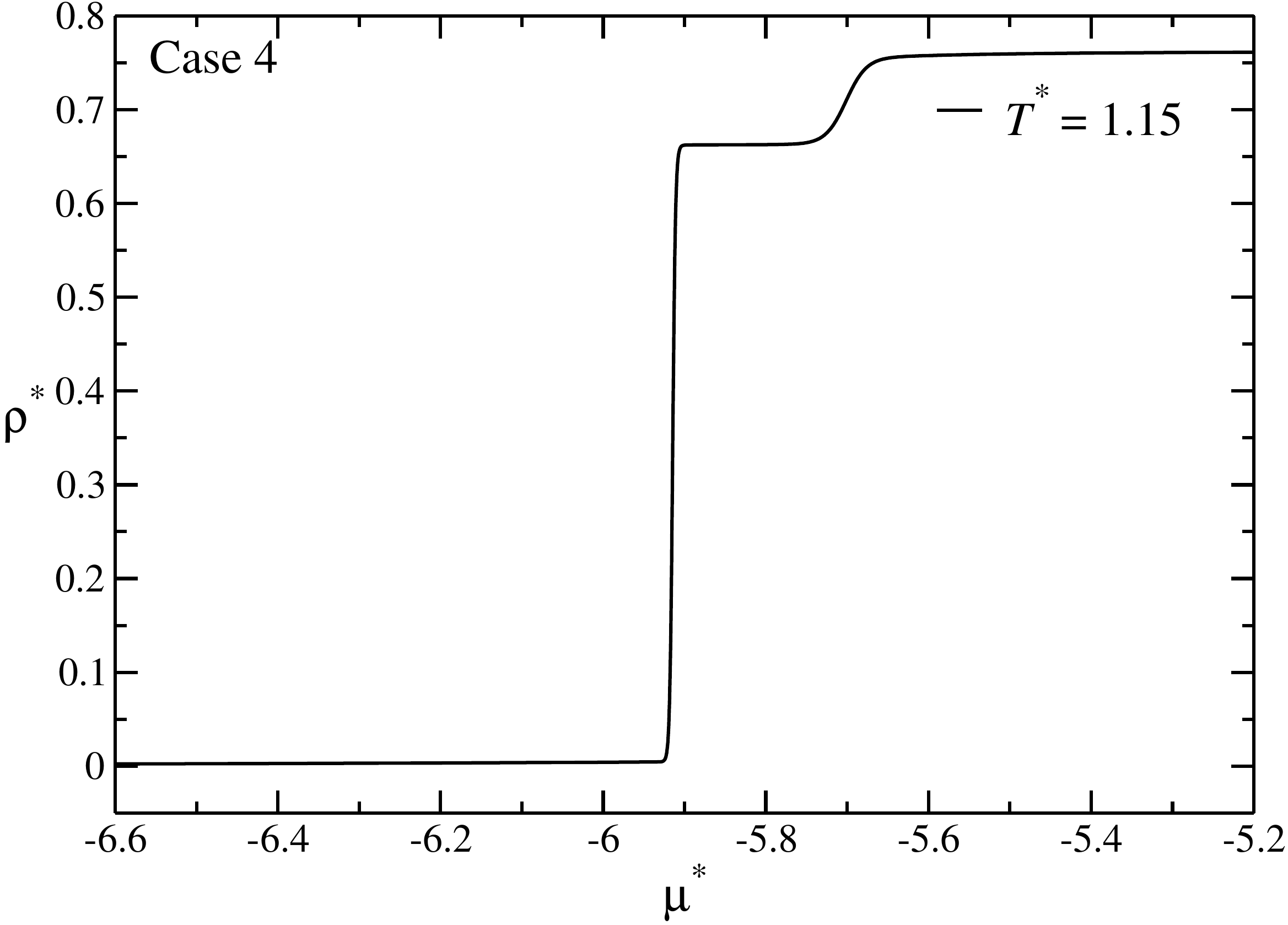}
\caption{Density versus chemical potential for the  case 2 at
  $T^{\ast}=0.1221$, case 3  at $T^{\ast}=0.58$ and case 4  at
  $T^{\ast}=1.15$. }
\label{fig:rhoxmus}
\end{figure}
%%%%%%%%%%%%%%%%%%%%%%%%%%%%%%%%%%%%%%%%

The nature of the liquid-gas phase transition is Ising-like
since the method we have used to characterize the gas-liquid critical point 
is based on
an Ising 3D universality class hypothesis.
The figure.~\ref{fig:crit_allB} illustrates the normalized 
probability distribution, $P(x)$,  which has an
universal form at criticality represented by
$x=A(M-M_c)$ where $A$ is the non-universal 
constant and the critical value of the
ordering operator $M_c$ were chosen so that 
the data have zero mean and unit variance.
For one-components systems
 the ordering operator, $M$, is
 proportional to a linear combination of the number
of particles $N$ and the system total configurational energy $U$, given by
%%%%%%%%%%%%%%%%%%%%%%%%%
\begin{equation}
 M\propto N-sU,
\end{equation}
%%%%%%%%%%%%%%%%%%%%%%%%%
where $s$ is the field-mixing parameter~\cite{Wilding}. This 
approach is appropriated only if the symmetry of the order
parameter is Ising-like.

The figure~\ref{fig:txrho_phase} also
shows for high densities a plateau that suggests the 
presence of a coexistence between two  high density
phases. Unfortunately these transition does
no appear clearly in the method we are employing, possible
for not being Ising-like.
In order to obtain some evidence of the 
existence of another type of transition in
addition to the liquid-gas, the behavior 
density versus chemical potential
was computed using Grand Canonical Monte
Carlo without the use of the Ising 3D universality class hypothesis.

The figure~\ref{fig:rhoxmus} illustrates the density versus
chemical potential for a fixed temperature showing 
a discontinuous change in density from the gas to the 
liquid and then between two liquids of high density.
The transition is first order.

A more clear picture of the nature of
the transition is obtained by 
computing 
the specific heat at constant volume. In the 
 grand canonical  simulations $c_V$ is obtained 
by using the expression~\cite{Al89}
%%%%%%%%%%%%%%%%%%%%%%%%%%%%%%%%%%%fig:txrho_phase
\begin{equation}
  \label{eq:cv}
   c_V=\frac{3}{2}k_B+\frac{1}{Nk_BT^2}\prt{
     \aver{\Delta U^2}_{\mu V T}-
     \dfrac{\aver{\Delta U \Delta N}_{\mu V T}^{2}}
     {\aver{\Delta N^2}_{\mu V T}}}\; .
\end{equation}
%%%%%%%%%%%%%%%%%%%%%%%%%%%%%%%%%%%

\begin{figure}[!htb]
\includegraphics[scale=0.35]{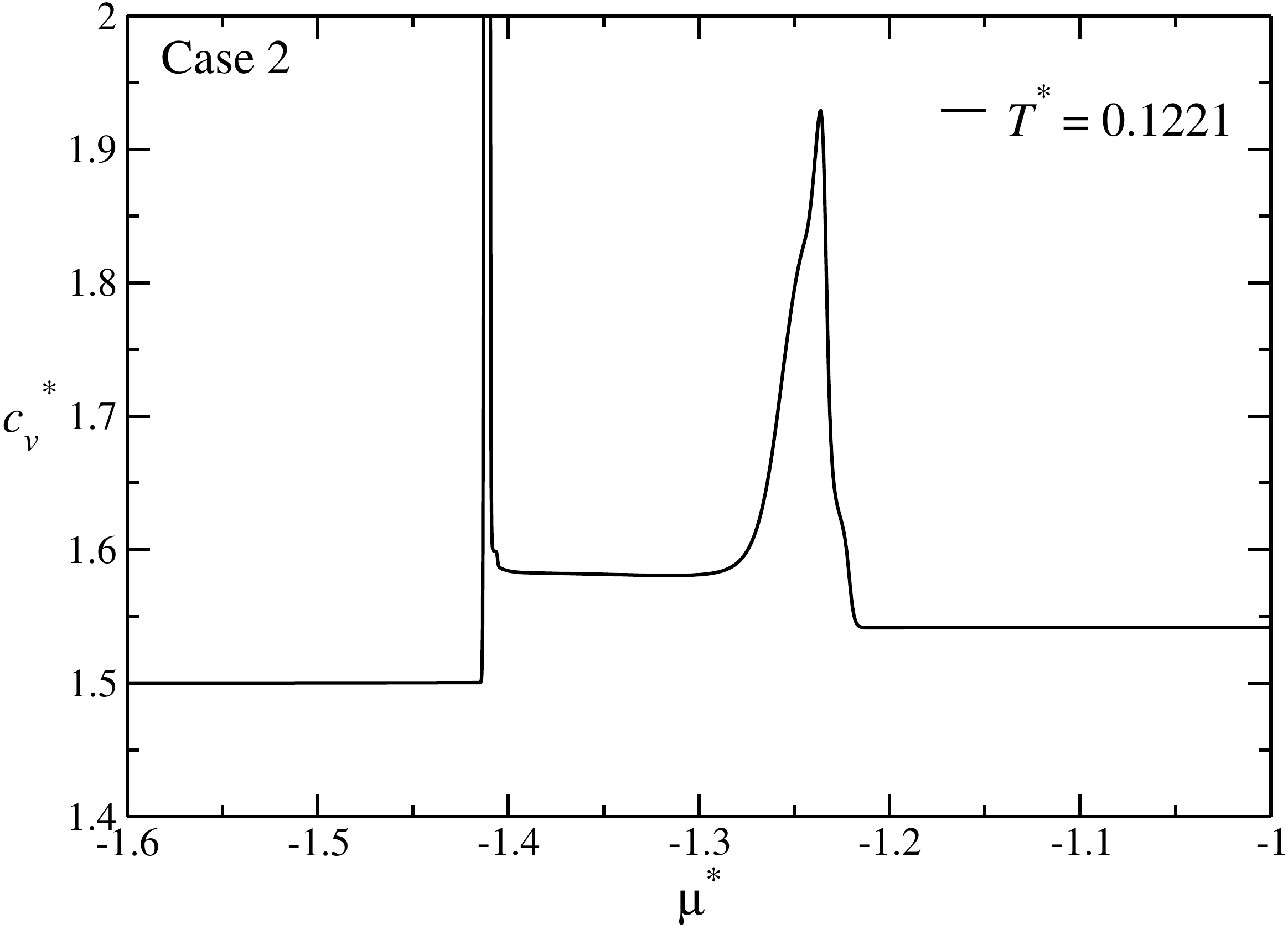}
\includegraphics[scale=0.35]{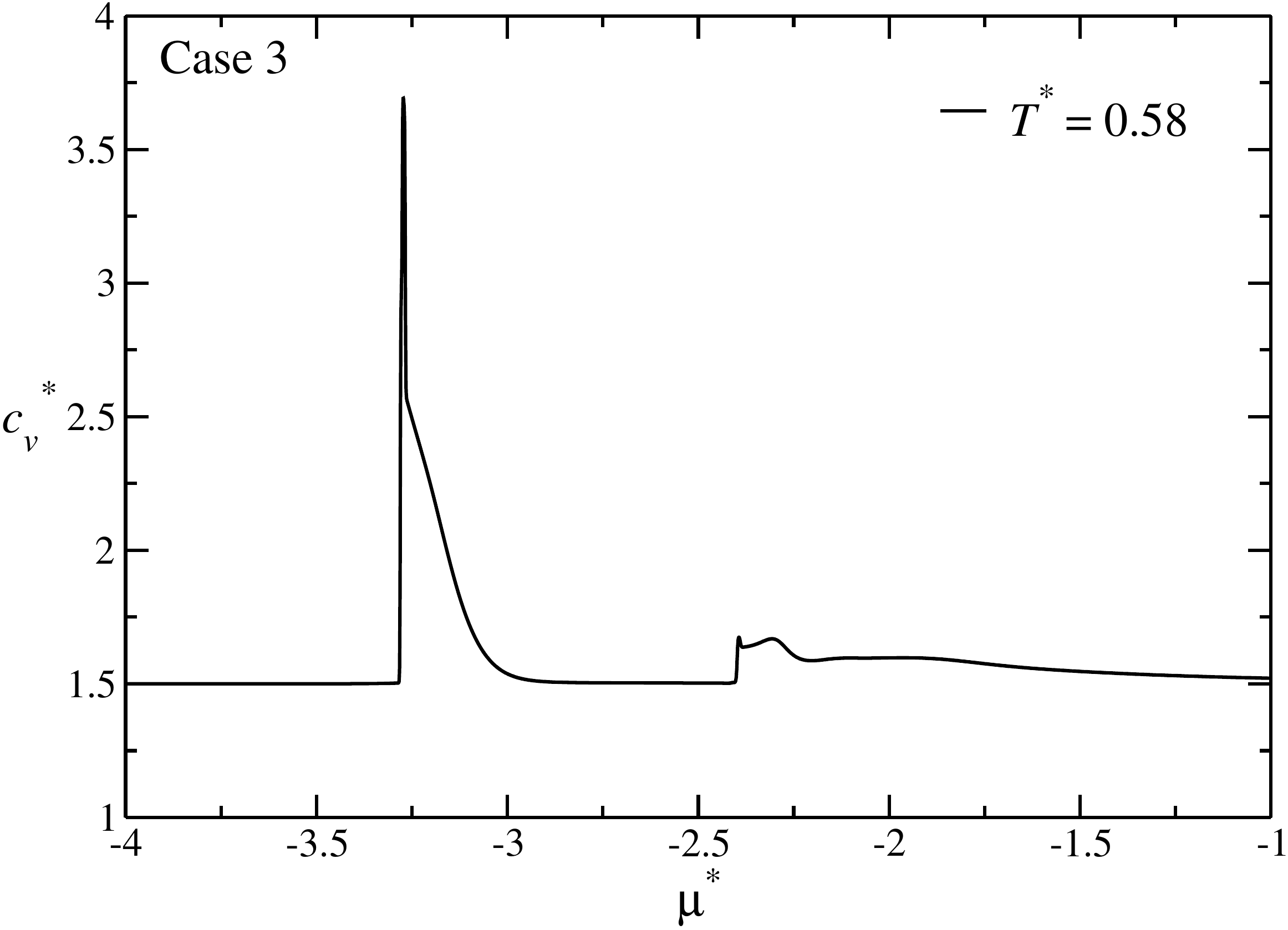}
\includegraphics[scale=0.35]{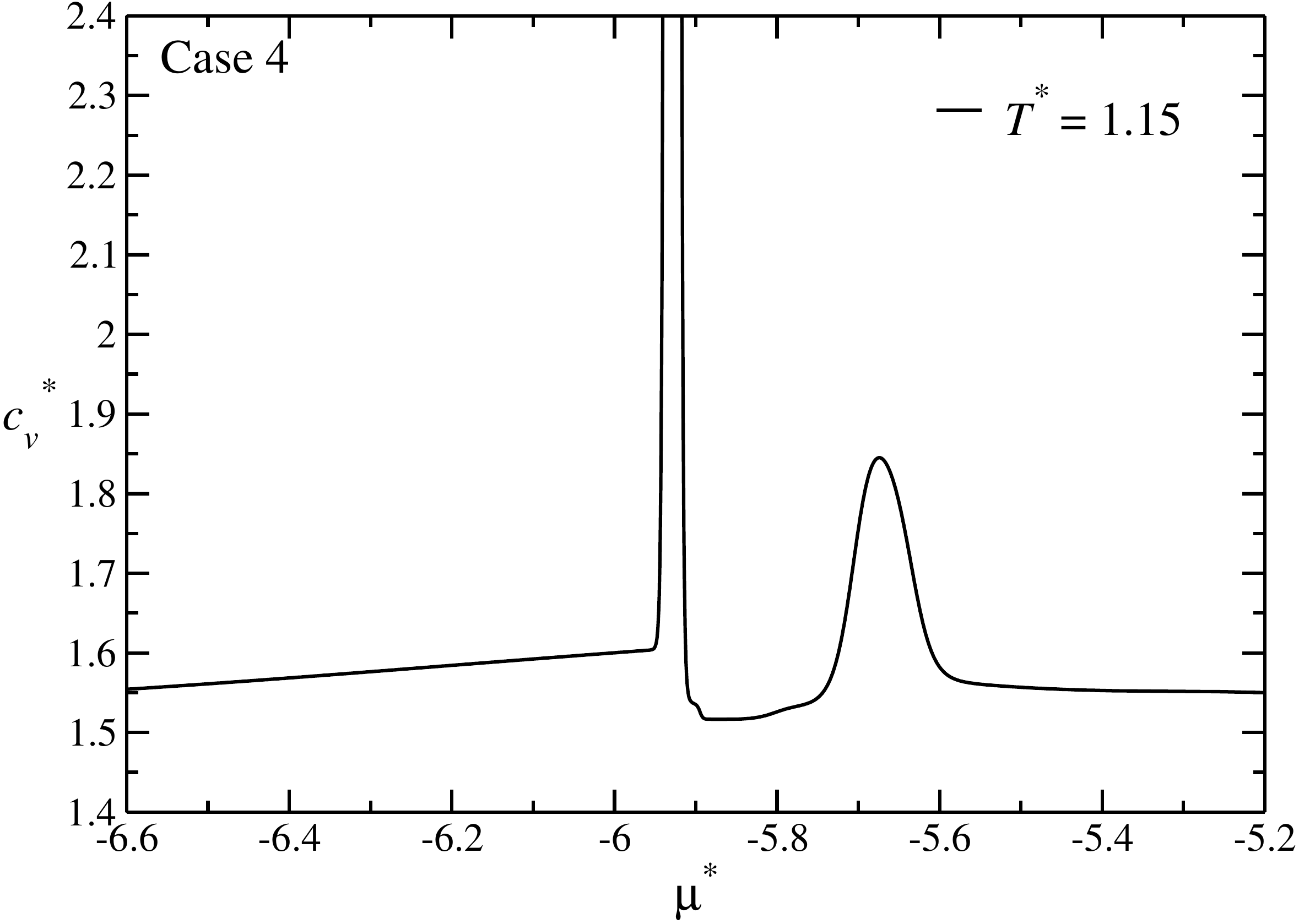}
\caption{Specific heat versus chemical
potential for the case 2 at
  $T^{\ast}=0.1221$, case 3  at $T^{\ast}=0.58$ and case 4  at
  $T^{\ast}=1.15$. }
\label{fig:cvxmus}
\end{figure}

The figure~\ref{fig:cvxmus} shows the 
behavior of the specific heat versus chemical
potential at constant temperature for the 
three cases in which the potential shows
an attractive part. The graphs
show a large peak in the specific heat
that coincides with the liquid-gas phase
transition and a small peak that coincides
with the plateau in the figure~\ref{fig:txrho_phase}.
As the temperature is increased the large peak gives
rise to the liquid-gas critical point but the
small peak vanishes. This result suggests that
the phase observed in this region is either
solid or amorphous. Unfortunately lower 
temperature analysis is not feasible due
to the slowing down.
In the Monte Carlo analysis no indication of the 
liquid-liquid phase transition is observed. This
might be due to the present of  stable solid
phases and to the universality class of the
liquid-liquid transition that might not be 3D Ising-like 
as implied in our method.

%%%%%%%%%%%%%%%%%%%%%%%%%%%%%%%%%%%%%%%%
%%%%%%%%%%%%%%%%%%%%%%%%%%%%%%%%%%%%%%%%

%%%%%%%%%%%%%%%%%%%%%%%%%%%%%%%
\section{Conclusions}
\label{sec:conclusion}
%%%%%%%%%%%%%%%%%%%%%%%%%%%%%%%

In this paper we analyzed the
effect of the attractive part in the pressure
versus temperature phase diagram of a family of smooth
 a core-softened potential with  two
length scales. The four potentials analyzed  show  a repulsive
 shoulder  followed by well that in three cases are  attractive. We
 found that the increase of the attractive scale 
adds a negative pressure which shrinks the 
  TMD  region and it moves it to
lower pressures.
Related to the shrink of the TMD pressure region,
the  increase of the attractive well makes the density
versus pressure at constant temperature non monotonic 
and reentrant what is the necessary condition 
for the appearance of the two liquid phases.

Since all the anomalies are interconnected the attraction 
also shrinks the pressure range of the structural order parameter,
diffusion anomaly and excess entropy. 

In order to understand the nature of the liquid-gas phase
transition of the core-softened potentials, the three
systems in which attraction is present were analyzed using  GCMC.
We show that the liquid-gas phase transition is Ising like 
and the critical temperature increases with the increase
of the attractive well in the potential. The Monte Carlo
analysis does not show the presence of the liquid-liquid phase
transition what can be attributed to two 
effects:  the method employed 
that assume that the transition is 3D Ising and 
the presence of solid or amorphous phases are present 
obscuring the high density liquid phase that might be metastable.

%%%%%%%%%%%%%%%%%%%%%%%%%%%%%%%
\section*{ACKNOWLEDGMENTS}
%%%%%%%%%%%%%%%%%%%%%%%%%%%%%%%
We tank for financial support the Brazilian science agencies CNPq,
CAPES and INCT-FCx and Centro de Física Computacional IF(CFCIF) for
computational support.

%%%%%%%%%%%%%%%%%%%%%%%%%%%%%%%%%%%%%%%%%%%%%%%%%%%%%%%%%%%%%

%\bibliographystyle{abnt}
\bibliography{Biblioteca}

\end{document}